\documentclass[twocolumn]{aastex63}

\usepackage{float}
\usepackage{amsmath,amssymb}
\usepackage{natbib}    
\usepackage{hyperref} 
\usepackage{graphicx} 
\usepackage{color}
\usepackage{verbatim}    
\usepackage{enumitem}
\usepackage{natbib}
\usepackage{CJK}
\usepackage{multirow}

\newcommand{\proptosim}{\mathrel{\vcenter{
 \offinterlineskip\halign{\hfil$##$\cr
 \propto\cr\noalign{\kern2pt}\sim\cr\noalign{\kern-2pt}}}}}
\renewcommand{\b}[1]{\boldsymbol{#1}}

\newcommand{\mean}[1]{\langle #1\rangle}

\renewcommand{\min}{\mathrm{min}}

\newcommand{\au}{\mathrm{AU}}
\newcommand{\cm}{\mathrm{cm}}
\newcommand{\m}{\mathrm{m}}

\newcommand{\g}{\mathrm{g}}
\newcommand{\G}{\mathrm{G}}
\newcommand{\K}{\mathrm{K}}  
\newcommand{\km}{\mathrm{km}}

\newcommand{\eV}{\mathrm{eV}}
\newcommand{\keV}{\mathrm{keV}}
\newcommand{\s}{\mathrm{s}}
\newcommand{\yr}{\mathrm{yr}}
\newcommand{\hr}{\mathrm{hr}}
\newcommand{\ang}{\ensuremath{\text{\AA}}}
\newcommand{\dyn}{\mathrm{dyn}}

\newcommand{\lya}{\text{Ly}\ensuremath{\alpha~}}

\newcommand{\B}{\mathrm{B}}
\newcommand{\kb}{k_\mathrm{B}}

\renewcommand{\d}{\mathrm{d}}
\newcommand{\e}{\mathrm{e}}

\renewcommand{\ion}[2]
{{\rm#1}\;\textsc{\MakeLowercase{#2}}}
\renewcommand{\roman}[1]{\textsc{\MakeLowercase{#1}}}

\newcommand{\eff}{\mathrm{eff}}
\newcommand{\los}{{\rm LoS}}

\newcommand{\dust}{\mathrm{dust}}

\newcommand{\p}{{\rm p}}
\newcommand{\J}{{\rm Jup}}

\newcommand{\ad}{\mathrm{ad}}   
\newcommand{\eq}{\mathrm{eq}}   
\newcommand{\rcb}{\mathrm{rcb}} 
\newcommand*\chem[1]{\ensuremath{\mathrm{#1}}}
\newcommand{\pos}[1]{\ensuremath{\mathrm{#1}^+}}

\newcommand{\ext}[1]{\ensuremath{\mathrm{#1}^*}}

\newcommand{\figdir}{Figures}

\newcommand{\HeIline}{\ion{He}{I} $10830~\ang$\ }

\begin{document}

\title{Metastable Helium Absorptions with 3D Hydrodynamics
  and\\ Self-Consistent Photochemistry I:\\
  WASP-69\lowercase{b}, Dimensionality, XUV Flux Level,
  Spectral Types, and Flares}

\author{Lile Wang$^{1}$, Fei Dai$^{2}$}

\footnotetext[1]{Center for Computational Astrophysics,
  Flatiron Institute, \\ New York, NY 10010;
  lwang@flatironinstitute.org}
\footnotetext[2]{Division of
  Geological and Planetary Sciences, \\
  California Institute of
  Technology, Pasadena, CA 91125} 

\begin{abstract}
  The metastable Helium (\ext{He}) lines near $10830~\ang$
  are ideal probes of atmospheric erosion--a common
  phenomenon of close-in exoplanet evolution. A handful of
  exoplanet observations yielded well-resolved \ext{He}
  absorption features in transits, yet they were mostly
  analyzed with 1D isothermal models prescribing mass-loss
  rates. This work devises 3D hydrodynamics co-evolved with
  ray-tracing radiative transfer and non-equilibrium
  thermochemsitry. Starting from the observed
  stellar/planetary properties with reasonable assumptions
  about the host's high energy irradiation, we predict from
  first principle the mass loss rate, the temperature and
  ionization profiles, and 3D outflow kinematics. Our
  simulations well reproduce the observed \ext{He} line
  profiles and light curves of WASP-69b. We further
  investigate the dependence of \ext{He} observables on
  simulation conditions and host radiation. The key findings
  are: (1) Simulations reveal a photoevaporative outflow
  ($\sim 0.5~M_{\oplus}~{\rm Gyr}^{-1}$) for WASP-69b
  without a prominent comet-like tail, consistent with the
  symmetric transit shape \citep{2020AJ....159..278V}. (2)
  3D simulations are mandatory for hydrodynamic features,
  including Coriolis force, advection, and kinematic line
  broadening. (3) EUV ($>13.6~\eV$) photons dominate
  photoevaporative outflows and populate \ext{He} via
  recombination; FUV is also detrimental by destroying
  \ext{He}; X-ray plays a secondary role. (4) K stars hit
  the sweet spot of EUV/FUV balance for \ext{He} line
  observation, while G and M stars are also worthy
  targets. (5) Stellar flars create characteristic responses
  in the \ext{He} line profiles.
\end{abstract}


\section{Introduction}
\label{sec:intro}
One of the most exciting discoveries in exoplanetary
sciences in recent years is that the radii of sub-Neptune
planets have a bimodal distribution \citep{Fulton}. The
prevailing explanation is the atmospheric erosion by either
photoevaporation \citep[e.g.][]{Owen2013,Owen} or
core-powered mass loss \citep{Ginzburg, 2019MNRAS.487...24G,
  2020MNRAS.493..792G}. In any case, the prominence of the
radius gap implies that atmospheric erosion is probably a
stage of evolution that close-in exoplanets very commonly go
through. Since the discovery of the Lyman $\alpha$ (\lya
hereafter) transit of hot Jupiter HD209458 b
\citep{2003Natur.422..143V}, \lya has been a workhorse for
studying atmospheric erosion \citep[e.g.][]
{2010A&A...514A..72L,2014ApJ...786..132K}. However, \lya has
some unavoidable limitations, namely, it is heavily
contaminated by geocoronal emission, and interstellar
absorption saturating the very center of the line
\citep[e.g.][]{Ehrenreich}. Moreover, one has to go space to
observe this UV transition. These effects significantly
limit the number of systems for which we can study \lya
transits.

Besides \lya, helium lines are emerging as promising outflow
indicators. The $2^3$S state of helium is often called the
``metastable state'' (\ext{He} hereafter), because the
$2^3$S$\rightarrow 1^1$S transition is a magnetic dipole
process with a slow spontaneous decay rate of
$A \simeq 1.3\times 10^{-3}~\s^{-1}$
\citep{1971PhRvA...3..908D, 2006sham.book..199D}. Meanwhile,
the transition between the lower $2^3$S and the upper
$2^3$P$_J$ states of helium consists of three lines with
$A > 10^7~\s^{-1}$, whose wavelengths in vacuum are
$10832.08~\ang$ (for the $J=0$ upper state), $10833.24~\ang$
($J=1$), and $10833.33~\ang$ ($J=2$) respectively. These
lines are often referred to as ``\HeIline lines'' or the
``metastable helium lines'', as they are radiatively
decoupled from the ground state. The abundance of helium,
the absence of geocoronal or intersteallar containmination,
the longevity of metastable state, and observability from
the ground together enabled the \ext{He} lines as an
excellent probe of ionized flows in various scenarios of
astrophysics, including quasars
\citep[e.g.][]{2011ApJ...728...94L}, stellar atmospheres and
outflows \citep[see][and references to the
article]{2003ApJ...599L..41E}, and T Tauri stars
\citep[][]{2007ApJ...657..897K}.

Over the years, researchers have proposed the \ext{He} lines
as a tracer of mass loss of close-in exoplanets
\citep{2000ApJ...537..916S, 2016MNRAS.458.3880T,
  2018ApJ...855L..11O}. It was the secure detection of
\citet{2018Natur.557...68S} that revitalized interest in
this unique transition. At the time of writing this paper,
several close-in exoplanets have transmission \ext{He} line
profiles resolved by ground-based spectrographs
\citep[e.g.][]{2018Sci...362.1384A, 2018Sci...362.1388N,
  2018A&A...620A..97S, 2020AJ....159..115K,
  2020ApJ...894...97N}. More recently,
\citet{2020AJ....159..278V} custom made a narrow band filter
specifically for the \ext{He} transitions on the
diffuser-based photometric system on Palomar/WIRC. The
resultant precise light curves of the \ext{He} is
complementary to the line profiles from the near-infradred
spectrographs. A lot of information about atmospheric
outflow is hiding in these line profiles and light curves
waiting to be interpreted. The models that are commonly used
in the literature to interpret these \ext{He} observations
are 1D spherical symmetric models that are isothermal
\citep{2018ApJ...855L..11O, 2019ApJ...881..133O,
  2020A&A...638A..61P}, or have prescribed heating
efficiency \citep{2020A&A...636A..13L}. The model is widely
recognized for its simplicity and effectiveness, however it
has to prescribe, rather than predict, a mass loss rate or a
temperature profile.
  
In this work, we build upon our previous model that conducts
hydrodynamics, self-consistent thermochemistry, and
ray-tracing radiative transfer to study the photoevaporation
of sub-Neptune planets \citep[][WD18
hereafter]{2018ApJ...860..175W}. We have streamlined the
code so that it is sufficiently fast to run in 3D to fully
capture outflow dynamics, and added various processes that
are relevant to the (de)population of \ext{He}. We will show
in this paper that using the observed stellar/planetary
properties and making reasonable estimate of the high energy
spectral energy distribution (SED) about the host star, our
model can predict mass loss rate, the temperature profile,
the ionization states, and synthesize the observed \ext{He}
line observables from first principles.

In this first work of a series, we focus on WASP-69b, which
is one of the first detections of \ext{He} line absorption
with well-resolved line profile
\citep{2018Sci...362.1388N}. Acknowledging the many
limitations of a 1D isothermal model,
\citet{2018Sci...362.1388N} did not tie their \ext{He} line
observation with a theoretical model. Instead they only
reported what the data showed directly. Notably,
\citet{2018Sci...362.1388N} reported an asymmetric transit
profile characterized by a longer-than-expected egress that
could be caused by a comet-like tail associated with the
mass loss. However, \citet{2020AJ....159..278V} suggests a
symmetric shape of transit using better-sampled light curves
with higher precision and signal-to-noise ratio
(SNR). Another interesting point about WASP-69b is the
apparent temporal variability of the \ext{He} transit depth
seen in \citet{2018Sci...362.1388N}. We will try to
understand these observations of WASP-69b in the framework
of our 3D hydrodynamic simulations. Afterwards, we will use
WASP-69 as a fiducial case to investigate the impact of
dimensionality, XUV flux level, and host spectral types on
the observables of the \ext{He} lines.

This paper is structured as follows. \S\ref{sec:methods}
describes our methods of numerical simulations and synthetic
observations. In \S\ref{sec:fiducial-models}, we present a
fiducial model of WASP-69b that well reproduces all current
observations. Based on this model,
\S\ref{sec:parameter-study} studies how various system
parameters impact the rate of photoevaporation and \ext{He}
observables. \S\ref{sec:parameter-flares} explores the
possibility that stellar flare may cause the observed
temporal variability of WASP-69b.  \S\ref{sec:summary}
summarizes the findings of this paper.

\section{Methods}
\label{sec:methods}

\subsection{Basic Setup}
\label{sec:method-setup}

We conceptually divided a planet into four regions: (1) a
dense core, (2) a convective inner atmosphere, (3) a
quasi-isothermal outer atmosphere with equilibrium
temperature $T_\eq$ and (4) an outflowing region irradiated
by high energy photons \citep[e.g.][]{2006ApJ...648..666R,
  2016ApJ...825...29G, 2016ApJ...817..107O}. The equilibrium
temperature $T_\eq$ satisfies,
\begin{equation}
  \label{eq:T_eq}
  T_\eq \simeq 886~\K
  \left(\dfrac{L_*}{L_\odot}\right)^{1/4}
  \left(\dfrac{a}{0.1~\au}\right)^{-1/2}\ ,
\end{equation}
where $L_*$ is the bolometric luminosity of the star, and
$a$ is the semi-major axis of the planetary orbit.  Our
simulations will focus on regions (3) and (4), whereas the
structure of regions (1) and (2) provide the correct
boundary conditions crucial for correctly reproducing the
measured mass and radius of the planet. We refer the reader
to Appendix~\ref{sec:appdx-internal} for the details of how
we set up the internal structures of our planet and
resultant boundary conditions for our simulations.

We characterize the high energy radiation spectral energy
distribution (SED) of the host star with 5 different
characteristic energy bins: (1) $h\nu = 2~\eV$ for infrared,
optical and near ultraviolet (NUV) photons, (2)
$h\nu = 7~\eV$ for ``soft'' far ultraviolet (FUV) photons
that can photoionize \ext{He}, (3) $h\nu = 12~\eV$ for the
Lyman-Werner band FUV photons that can photodissociate
molecular hydrogen, (4) $h\nu = 20~\eV$ for ``soft'' extreme
ultraviolet (soft EUV) photons that can ionize hydrogen but
{\it not} helium, (5) $h\nu = 40~\eV$ for hard EUV photons
that ionize hydrogen {\it and} helium, and (6)
$h\nu = 3~\keV$ for the X-ray.

Our simulation combines ray-tracing radiative transfer,
real-time non-equilibrium thermochemistry, and full
hydrodynamics calculations (based on a higher order Godunov
hydrodynamic solver \verb|Athena++|;
\citealt{2020arXiv200506651S}). The simulation is mostly
based on our WD18 work with a few modifications and
improvements added for the higher dimensionality and the
inclusion of \ext{He}.

\subsection{Geometry and Boundary Conditions}
\label{sec:method-geo-kin}

The density distribution, temperature profile and the
dynamics of outflowing atmosphere all play a part in the
\ext{He} observables. To capture the outflow dynamics
accurately, simulations should include the gravity of the
star and the planet and the effects of orbital motion: the
centrifugal and Coriolis forces. Therefore, 3D simulations
are required. Given its short orbital period and observed
radial velocities \citep{2014MNRAS.445.1114A}, we assume
that WASP-69b is tidally locked and circularized.  Our
simulation is run in a co-rotating frame centered on the
planet. We adopt a spherical polar coordinate
$(r,\theta,\phi)$ with $\theta = 0$ pointing towards the
host star and $\phi= 0$ pointing in the direction of orbital
motion.

The mesh covers the domain
$(r,\theta,\phi) \in [r_{\rm in}, r_{\rm out}] \otimes
[0,\pi] \otimes [0,\pi]$. Planet-specific radial boundaries
$r_{\rm in}$ and $r_{\rm out}$ usually extend from the the
base of the quasi-isothermal layer to a relatively large
radii (150 $R_\oplus$ in this case) such that the
density/opacity drops low enough. The radial grids are
placed logarithmically to strategically capture the steep
change of density, while latitudinal and azimuthal grids are
spaced evenly. Reflecting boundary conditions are enforced
at the $r = r_{\rm in}$, $\phi = 0$ and $\phi = \pi$
boundaries, while the $r= r_{\rm out}$ boundary is an
outflowing boundary. The $\theta = 0$, $\theta = \pi$
boundaries are polar wedges to avoid coordinate
singularity. The whole mesh, with its polar axis always
pointing towards the host star, co-rotates with the orbital
motion and the rotation of the tidally-locked planet.

In a 3D spherical polar mesh, the grids near the polar axis
are narrow in the azimuthal direction
($\delta x_\phi \simeq r_{\rm cc} \sin \theta_{\rm cc}
\delta\phi$; subscripts ``cc'' stand for ``cell center'').
This can result in highly non-unitary aspect ratios
($\delta x_\theta \simeq r_{\rm cc} \delta\theta$), and a
stringent Courant-Friedrichs-Lewy (CFL) condition. We thus
introduce an adaptive ``mesh coarsening'' technique for the
azimuthal grids near the poles. Without any violations of
conservation laws, the effective aspect ratio of the
high-latitude zones become close to one and the timestep
constraints imposed by CFL condition is not as severe
\citep[see also][] {2019PASJ...71...98N,
  2019MNRAS.484.3307M}. This helps to greatly speed up our
model.

\subsection{\ext{He} in non-LTE Thermochemistry}
\label{sec:method-chem}

Our simulation includes a
non-local-thermodynamic-equilibrium reaction network that
coevolves with the hydrodynamics (see WD18 for detail). With
the addition of the metastable state of neutral helium and
all relevant reactions that populate and de-populate this
state (see \citealt{2018ApJ...855L..11O} and references
therein). Our reaction network now has 26 thermochemical
``species'' (24 chemical species in WD18, \ext{He}, and
internal energy density) and 135 reactions such as
ionization, recombination, collisional (de-)excitation,
photodissociation, and cooling and heating processes. The
ordinary differential equations (ODEs) of the thermochemical
network were solved efficiently using the semi-implicit
method specially optimized for the graphics processing units
(GPUs). The resultant efficiency allows us to coevolve the
hydrodynamics with the thermodynamics, rather than including
thermodynamics as a post-processing step that is often done
in the literature. Again, we refer interested readers to
WD18 for more details.

\subsection{Synthetic Observations}
\label{sec:method-spec}

We synthesize both the transmission line profiles
\citep{2018Sci...362.1388N} and the narrow-band light curves
\citep{2020AJ....159..278V} of \ext{He} transitions using
our simulations. At each wavelength $\lambda$ and a
particular orbital phase, the optical depth along a line of
sight (LoS hereafter) is given by,
\begin{equation}
  \label{eq:method-tau-los}
  \begin{split}
    \tau(\lambda; \Phi) = \int & |\d \mathbf{x}|\
    n(\ext{He};\mathbf{x};\Phi)
    \\
    & \times \sum_i \sigma_i[\lambda;\hat{n}_\los\cdot
    \mathbf{v}(\mathbf{x};\Phi),
    T(\mathbf{x};\Phi)]\ .
  \end{split}
\end{equation}
where we have transformed from our planet-centered
coordinate systems in the simulations to a star-centered
coordinate system for the synthetic observations. Thus
$\mathbf{x}$ and $\mathbf{v}$ are the position and velocity
vector measured from the host star. The integration goes
along the designated LoS, and the summation index $i$ runs
over \ext{He}'s three lines with different upper state
quantum number $J$.  The cross-section $\sigma_i$ is assumed
to be a Voigt profile which convolves the intrinsic Lorentz
profile ($\gamma = A / 4\pi$,
$A = 1.0216\times 10^7\ \s^{-1}$; see
\citealt{2006sham.book..199D}) with a Gaussian profile from
thermal broadening at temperature $T(\mathbf{x})$. This
Voigt profile is shifted by the local bulk velocity
including orbital motion and outflow kinematics and the
projection onto the LoS
$\hat{n}_\los\cdot \mathbf{v}(\mathbf{x})$.

This integration is performed for all relevant LoS that
originate from the surface of the host star:
\begin{equation}
  \label{eq:method-extinction}
  \epsilon(\lambda;\Phi) = 1 -
  \int \d\Sigma\ S \e^{-\tau(\lambda;\Phi)}\ ,
\end{equation}
where $\epsilon(\lambda)$ is the relative extinction at
wavelength $\lambda$, $S(\lambda)$ is the normalized surface
brightness ($\int\d\Sigma S(\lambda) = 1$) of the star after
accounting for limb darkening and stellar rotation. The
integral runs through the entire projected
stellar surface.


$\epsilon(\lambda;\Phi)$ is the absorption line profile as a
function of wavelength and orbital phase (time). We mimic
what observers often do in \ext{He} observations i.e. time
averaging $\epsilon(\lambda;\Phi)$ over the entire transit
event from nominal ingress to egress ($t_\roman{ii}$ through
$t_\roman{iii}$)\footnote{Following the conventions, in this
  paper, we use $t_\roman{i}$ and $t_\roman{ii}$ for the
  start/end of the ingress, and $t_\roman{iii}$ and
  $t_\roman{iv}$ for those of the egress.}. The outcome
$\Delta\epsilon(\lambda)$ is a line profile of excess
absorption to be compared with observations directly.

We report a number of summary statistics including the
equivalent widths
$\mean{W_\lambda} \equiv \int\Delta
\epsilon(\lambda)~\d\lambda$, the radial velocity shift of
the absorption peak $\Delta v_{\rm peak}$ and the
full-width-half-maximum (FWHM) of the absorption line
profile. These summary statistics help us to compare between
models and observations efficiently and are less prone to
measurement uncertainty, bad pixels and other instrumental
effects.

Finally, we integrated $\epsilon(\lambda;\Phi)$ multiplied
by a filter bandpass function over $\lambda$. The result is
a transit light curve near the \ext{He} transitions. In this
work, we use the bandpass function provided by
\citet{2020AJ....159..278V} for a direct comparison with
their results.

\section{Fiducial Model of WASP-69\lowercase{b}}
\label{sec:fiducial-models}

\begin{deluxetable}{lr}
  \tablecolumns{2} 
  \tabletypesize{\scriptsize}
  \tablewidth{0pt}
  \tablecaption
  {Properties of the fiducial model for WASP-69b
    \label{table:wasp69b-fiducial} }   
  \tablehead{
    \colhead{Item} &
    \colhead{Value}
  }
  \startdata
  Simulation domain & \\
  Radial range & $11.37 \le (r/R_\oplus) \le \ 150$\\
  Latitudinal range & $0\le\theta\le\pi$ \\
  Azimuthal range & $0\le\phi\le\pi$ \\
  Resolution $(N_{\log r}\times N_{\theta} \times N_{\phi})$ &
  $144\times 128\times 64$ \\ 
  \\
  Planet interior$^\dagger$ & \\
  $M_\rcb$ & $82.6~M_\oplus$ \\
  $\mean{r_\eff}$ & $11.8~R_\oplus$ \\
  \\
  Radiation flux [photon~$\cm~^{-2}~\s^{-1}$] & \\
  $2~\eV$ (IR/optical)  & $6.4\times 10^{19}$ \\
  $7~\eV$ (Soft FUV)   & $6\times 10^{15}$ \\
  $12~\eV$ (LW)   & $1\times 10^{12}$ \\
  $20~\eV$ (Soft EUV) ${}^\ddagger$  & $5\times 10^{12}$ \\
  $40~\eV$ (Hard EUV) ${}^\ddagger$  & $3\times 10^{13}$ \\
  $3~\keV$ (X-ray) ${}^\ddagger$  & $1.2\times 10^{12}$ \\  
  \\
  Initial abundances [$n_{\chem{X}}/n_{\chem{H}}$] & \\
  \chem{H_2} & 0.5\\
  He & 0.1\\
  \chem{H_2O} & $1.8 \times 10^{-4}$\\
  CO & $1.4 \times 10^{-4}$\\
  S  & $2.8 \times 10^{-5}$\\
  Si & $1.7 \times 10^{-6}$\\
  Gr & $1.0 \times 10^{-7}$ \\
  \\
  Dust/PAH properties & \\
  $\sigma_\dust/\chem{H}$ (Effective specific cross section)
  & $8\times 10^{-22}~\cm^2$
  \enddata
  \tablecomments{
    $\dagger$: See Appendix~\ref{sec:appdx-internal};
    Thomson opacity is used for $\mean{r_\eff}$.\\
    $\ddagger$: The inferred values of fluxes in
    \citet{2018Sci...362.1388N} for WASP-69b are
    $2.6\times 10^{13}~{\rm ph\ cm}^{-2}{\rm s}^{-1}$ for
    EUV (represented by $h\nu = 40~\eV$ photons) and
    $0.5\times 10^{12}~{\rm ph\ cm}^{-2}{\rm s}^{-1}$ for
    X-ray (represented by $h\nu = 3~\keV$ photons)
  }
\end{deluxetable}
In this section, we will show how we arrived at a fiducial
model that gives rather remarkable agreement with the
observed \ext{He} line profiles \citep{2018Sci...362.1388N}
and light curves \citep{2020AJ....159..278V}. We note that
our 3D hydrodynamic model is not fast enough \footnote{Even
  with a GPU-accelerated infrastructures, each simulation
  takes about 5 hours on one computation node with 40 CPU
  cores (Intel Skylake) and 4 GPUs (Nvidia Tesla V100) on
  the Popeye-Simons Computing Cluster.} for a full
exploration of the parameter space with techniques such
Markov Chain Monte Carlo or even simple gradient
descent. Instead we had to rely on the reported system
parameters and making reasonable assumption as well as hand
tuning the high energy SED of the host star. We will see
shortly, without much tuning, we can arrive at a fiducial
model that fits various observations of WASP-69b very well.
  
We set up our simulations to match the reported system
properties of WASP-69b \citep{2014MNRAS.445.1114A}. The host
star is K star with $M_* = 0.826~M_\odot$,
$R_* = 0.813~R_\odot$ and $T_{\eff} = 4715~\K$. WASP-69b has
a circular orbit with semi-major axis $a = 0.04525~\au$.
The equilibrium temperature is estimated using
Eqn. \ref{eq:T_eq} $T_\eq = 965~\K$. The planet has an
optical transit radius of $R_\p \simeq 1.057~R_\J$ and a
mass of $M_\p \simeq 0.26~M_\J$ from radial velocity
follow-ups. Details of the fiducial model are presented in
Table~\ref{table:wasp69b-fiducial}.

The interior of our WASP-69b model is set up as described in
Appendix~\ref{sec:appdx-internal}. The core size, the
equation of state and other details of the interior of a
giant planet is still subject to a lot of uncertainties even
in the case of Jupiter \citep[see
e.g.][]{2017GeoRL..44.4649W}. However, the details of the
interior should not affect the outflowing region of the
envelope which is what we are interested in this work. We
set the inner boundary of our simulation at $11.37~R_\oplus$
so that we capture several scale heights below the optical
transit radius $\mean{r_\eff}\simeq R_\p$ at
$1.057~ R_{\rm Jup}$. The outer boundary is located at
$150~R_\oplus$. For simplicity, we assumed an atmospheric
metallicity often seen in protoplanetary disk (WD18) which
is slightly below solar value
(Table~\ref{table:wasp69b-fiducial}). We will explore any
metallicity dependence in a future work.

Optical and infrared fluxes, represented by the
$h\nu = 2~\eV$ photon energy bin, are calculated using the
host star radius and effective temperature. For the high
energy photons, more uncertainties are involved depending on
the age/activity of the host star; while direct measurements
are also lacking. We note that WASP-69 is moderately active
indicated by the \ion{Ca}{II} H and K lines
$\log\;R^\prime_{\rm H,K}=-4.54$
\citep{2014MNRAS.445.1114A}. As we will see later in \S
\ref{sec:XUV}, the \ext{He} absorption line profile depends
critically on the shares taken by various high-energy
radiation bins. After gaining intuition on how each energy
bin affects the \ext{He} line profiles (again \S
\ref{sec:XUV}), we varied the high energy SED of WASP-69
until we achieved a reasonable agreement with both the line
profile and light curve measurements. The resultant high
energy SED is quite typical of a K5 star when compared to
observational constraint of FUV and EUV flux of the MUSCLES
survey \citep{2016ApJ...820...89F, 2016ApJ...824..101Y,
  2016ApJ...824..102L, 2017ApJ...843...31Y}, the X-ray flux
according to \citet{1992A&A...264L..31G}, and the more
comprehensive compilation of \citet{2019ApJ...881..133O}.

\subsection{A Photoevaporative Outflow on WASP-69b}

\begin{figure*}
  \centering
  \hspace*{-0.2in}    
  \includegraphics[width=7.5in, keepaspectratio]
  {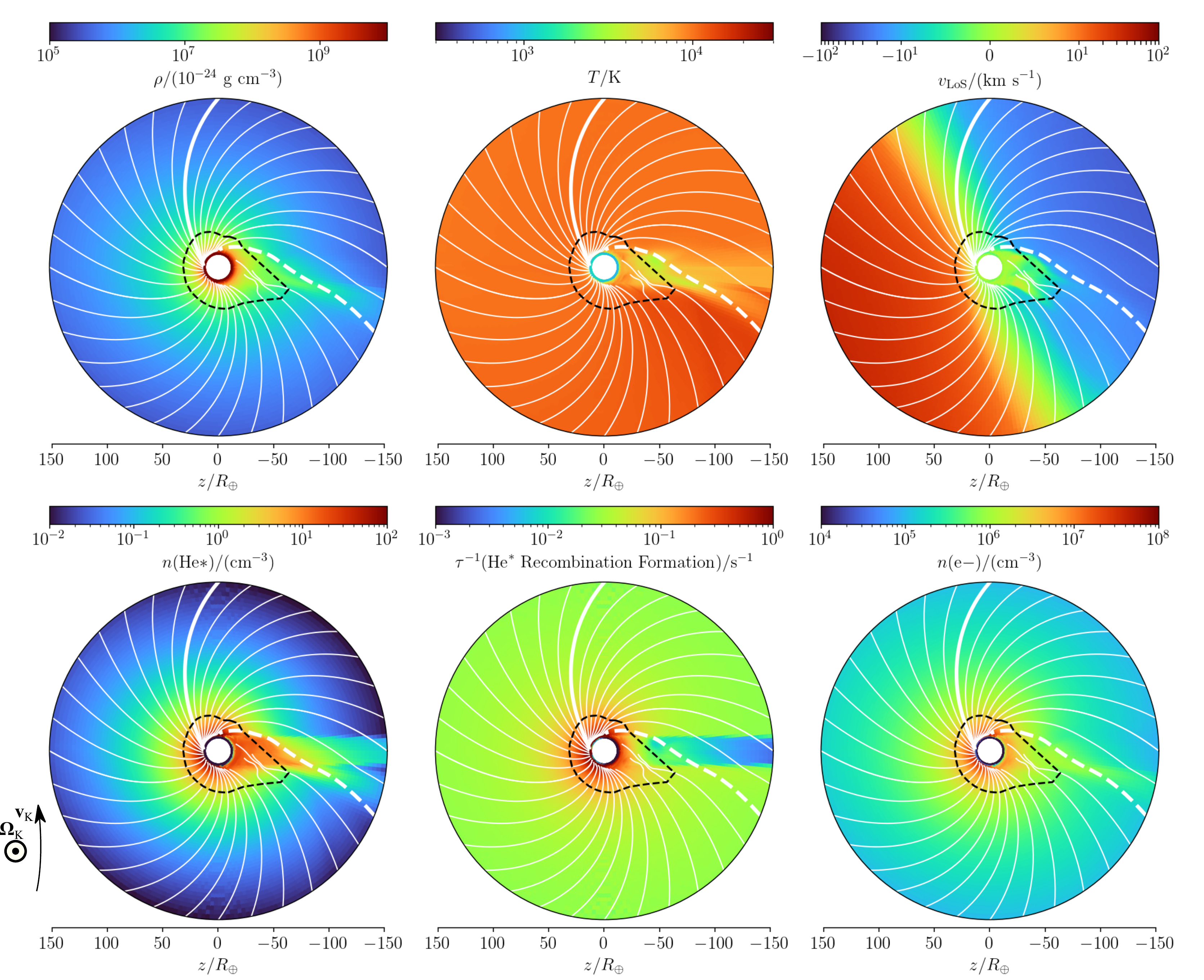}
  \caption{Profiles of the simulation for WASP-69b (fiducial
    model 69-0) in its quasi-steady-state. Stellar radiation
    comes from the left of the plot, and the orbital angular
    velocity $\mathbf{\Omega}_\K$ points out of the paper
    plane; the Keplerian motion of the plaent is upwards
    $\mathbf{v}_\K$. Colormaps show the mass density $\rho$
    ({\bf upper left panel}), temperature $T$ ({\bf upper
      middle}), line-of-sight velocity $v_\los$ ({\bf upper
      right}; the value is measured at mid-transit),
    \ext{He} number density $n(\ext{He})$ ({\bf lower
      left}), inverse timescale of recombination \ext{He}
    formation (defined as formation rate normalized by
    $n(\ext{He})$; {\bf lower middle}), and free electron
    number density $n(\e^-)$ ({\bf lower right}). White
    streamlines (projected to the orbital plane) are
    overlaid on each panel; two neighbor streamlines are
    separated in such a way that they are
    $\Delta \theta = \pi / 16$ apart when they reach the
    outer radial boundary ($r = 150~R_\oplus$). The heavy
    streamline are the reference lines on which the profiles
    are plotted in Figure~\ref{fig:wasp69b-profile}). Black
    solid lines indicate the sonic surface.}
  \label{fig:wasp69b-slice} 
\end{figure*}

\begin{figure*}
  \centering
  \includegraphics[width=7.0in, keepaspectratio]
  {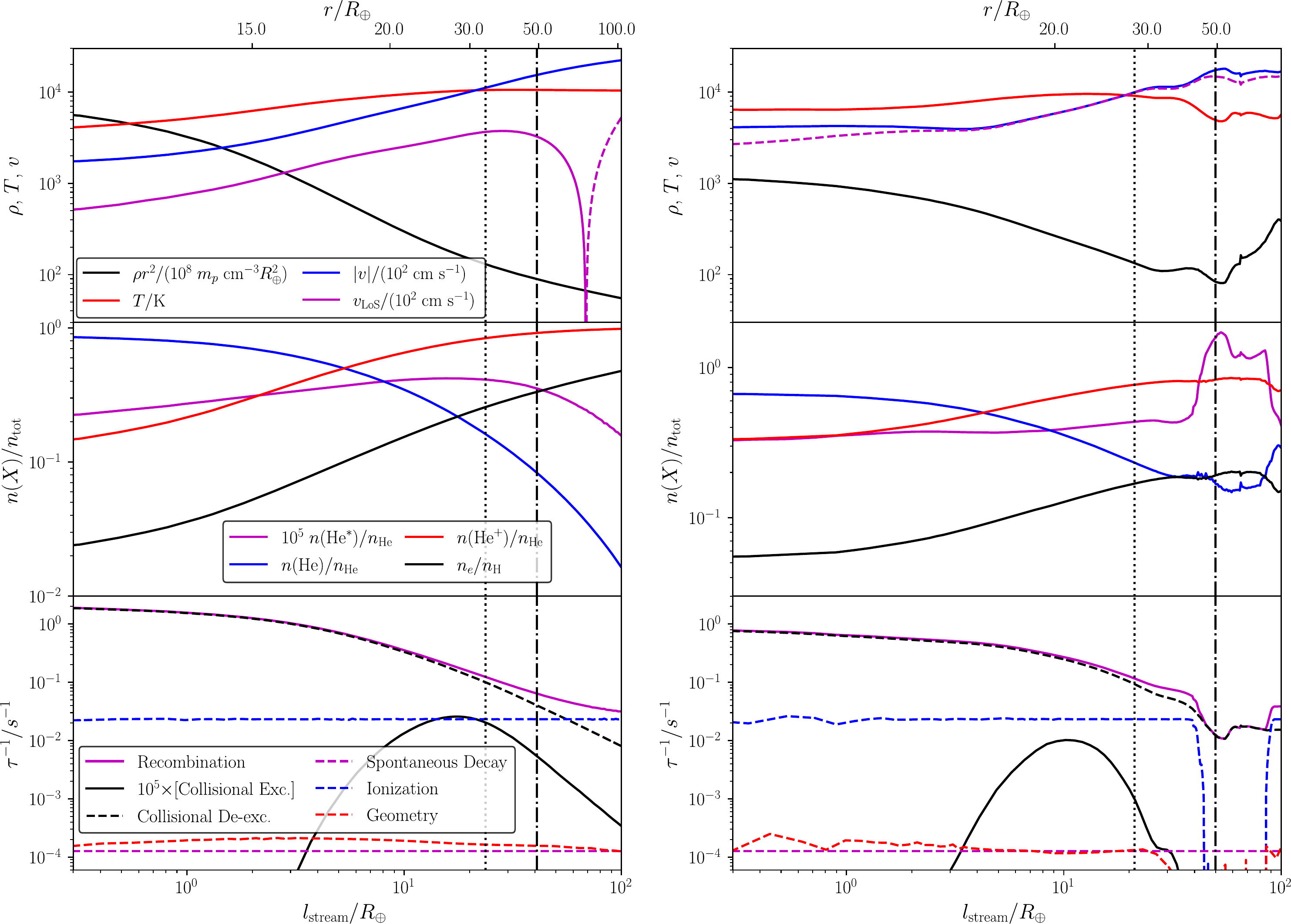}
  \caption{Key quantities of our fiducial model for WASP-69b
    (Model 69-0) along the two reference streamlines
    (plotted as the heavy streamlines in
    Figure~\ref{fig:wasp69b-slice}). The top abscissa is the
    radial coordinate $r$ corresponding to the curve length
    along the streamline on the bottom abscissa. {\bf The
      top panels} contain the profiles of scaled mass
    density $\rho r^2$, temperature $T$, velocity magnitude
    $|v|$ and the line-of-sight velocity $v_\los$. Dashed
    part of the $v_\los$ curve indicate negative
    values. {\bf The middle panels} present the abundances
    of free electrons (relative to total hydrogen nucleus
    density $n_{\rm H}$) and helium in different forms
    (relative to the total helium nucleus density
    $n_{\rm He}$; note that $n(\ext{He})$ is multiplied by
    $10^5$ for clarity). Inverse timescales of $\ext{He}$
    formation (in solid curves; note that the collisional
    excitation rate is multiplied by $10^5$) and destruction
    (in dashed curves) processes are shown in {\bf the
      bottom panels}. In all panels, the vertical dotted
    line indicates the sonic critical point, and the
    vertical dash-dotted line shows the location of Roche
    radius.  }
  \label{fig:wasp69b-profile}
\end{figure*}

\begin{figure}
  \centering
  \includegraphics[width=3.4in, keepaspectratio]
  {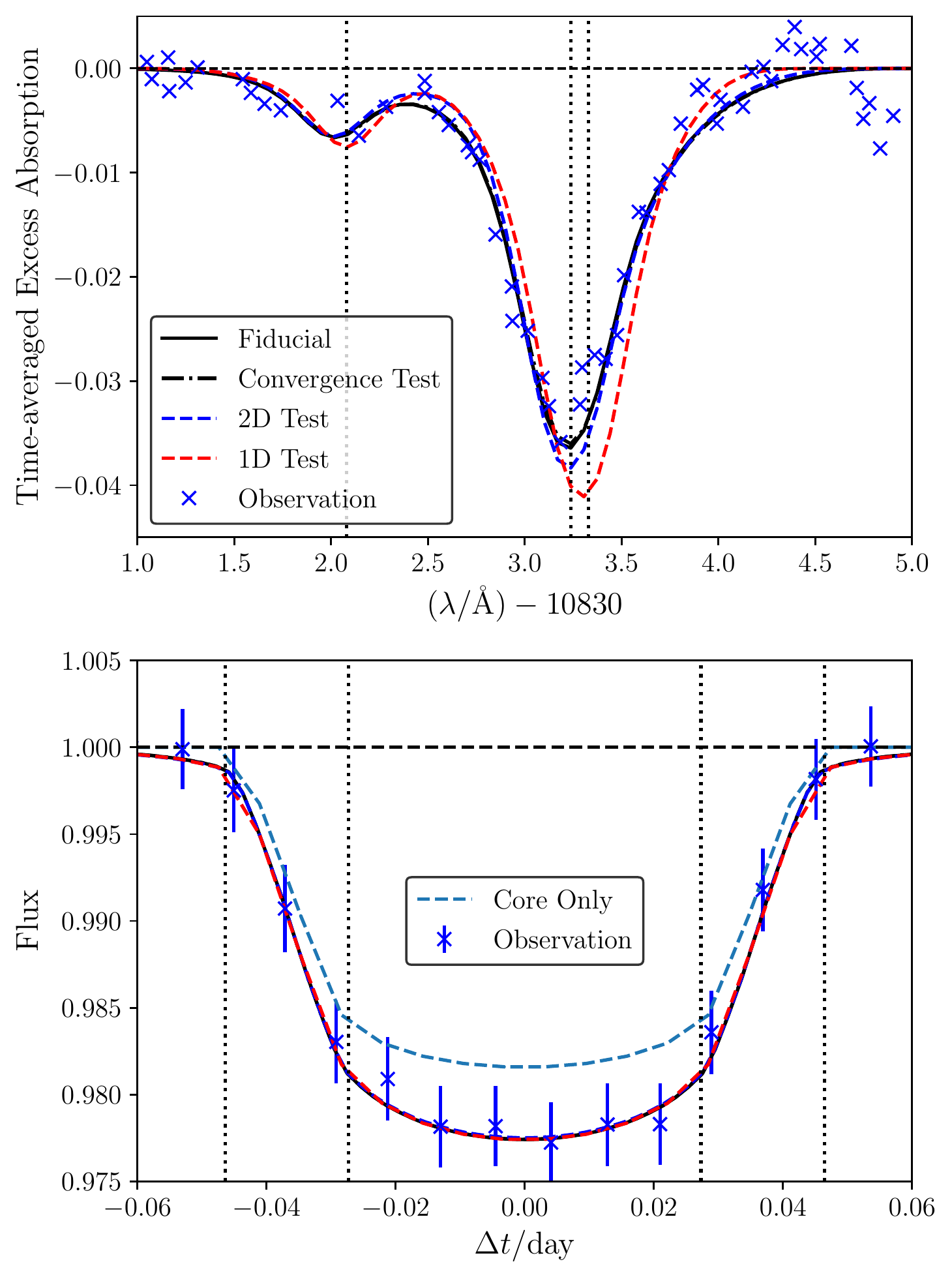}
  \caption{The observed and synthesized line profiles and
    light curves for WASP-69b. We include the results of our
    fiducial model (3D), 2D test, 1D test and a convergence
    test (same as fiducial but much higher grid resolution)
    (\S\ref{sec:method-setup}) as different line
    styles. Note that the curves for some test models
    overlap the fiducial curve and cannot be
    distinguished. {\bf The upper panel} presents the
    time-averaged excess absorption transmission spectra or
    the resolved line profile
    (eq.~\ref{eq:method-extinction}; time-averaged from the
    end of ingress through the begin of egress). Three
    vertical dotted lines indicate the three central
    wavelengths of \ext{He} triplet. {\bf The lower panel}
    compares the synthesized light curve with the
    observations and a hypothetical planet that {\it does
      not} have any atmospheres. We have rebinned the light
    curve data in \citet{2018Sci...362.1388N}.  The four
    vertical dotted lines indicate the ingress and egress.}
  \label{fig:wasp69b-spec}
\end{figure}

\begin{figure*}
  \centering
  \includegraphics[width=7.2in, keepaspectratio]
  {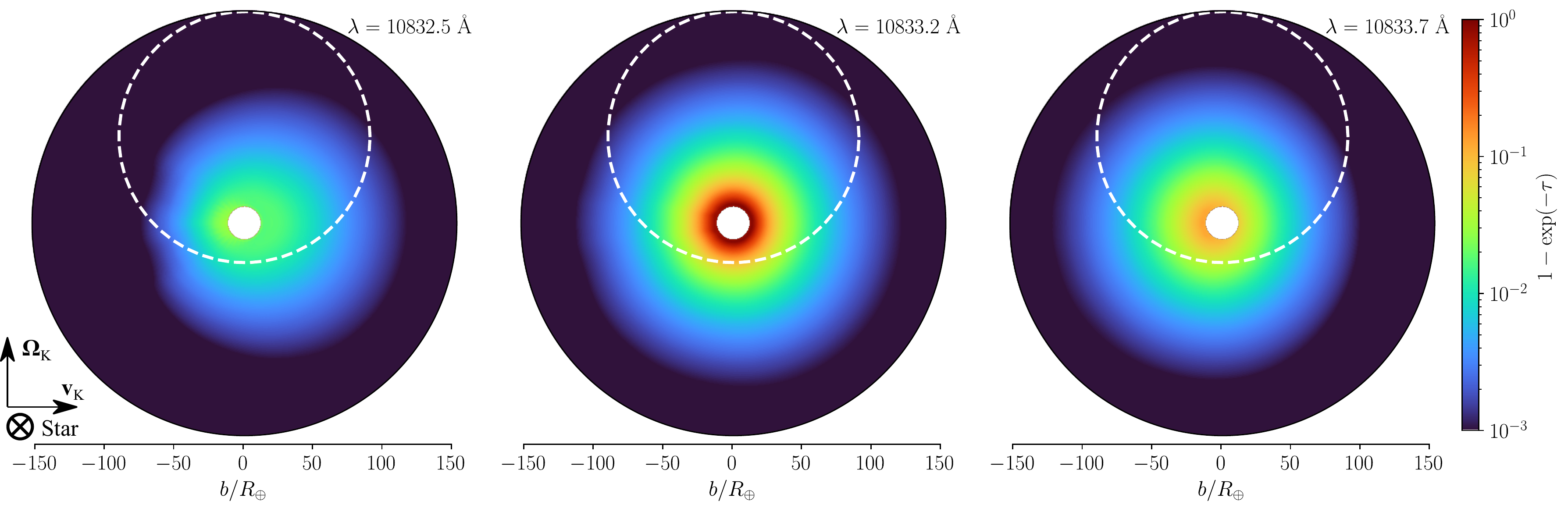}
  \caption{Extinction $[1-\exp(-\tau)]$ for Model 69-0 at
    three representative wavelengths
    $(\lambda/\ang)\in \{10832.5, 10833.2, 10833.7\}$ near
    the \ext{He} transitions. The profiles are measured at
    mid-transit (viz., all LoS are parallel to the
    star-planet line) and in a frame centered around the
    planet. The white dashed circles indicate the projected
    host star (with reported stellar radius and orbital
    inclination). The directions of orbit motion
    ($\mathbf{v}_\K$), orbital angular momentum vector
    ($\mathbf{\Omega}_\K$), and the direction to the host
    star are indicated at the lower-left corner of the whole
    figure.}
  \label{fig:wasp69b-ext}
\end{figure*}

Before analyzing our simulations, we ensured that
quasi-steady states have been achieved. This usually
involved running the simulations for many dynamical
timescales, specifically we set
$t_{\rm sim} \gtrsim 30~\tau_\dyn$. The dynamical timescale
$\tau_\dyn$ is estimated by the sound crossing time of the
Bondi radius:
\begin{equation}
  \label{eq:t-dyn}
  \tau_\dyn 
  \sim \dfrac{G M_\p}{c_s^3} \sim 1.2\times 10^3~\s\times
  \left(\dfrac{M_p}{10~M_\oplus}\right)
  \left(\dfrac{T}{10^4~\K}\right)^{-3/2}\ .
\end{equation}
Here, $c_s$ is the sound speed. For a typical $T=10^4~\K$
outflow we see for WASP-69b, $\tau_\dyn\sim
10^4~\s$. Moreover, we also check explicitly if the
simulation has settled down to a quasi-steady state by
comparing key hydrodynamic/thermodynamic properties in
neighboring dump files.

Our fiducial model for WASP-69b shows clear signs of a
photoevaporative outflow. In Figure~\ref{fig:wasp69b-slice},
we show 2D slices of the density, temperature and LOS
velocity distributions of our 3D simulations looking down
the North pole of the planet. We see a $T\gtrsim 10^4~\K$
hot ionized supersonic outflow originating at a wind base of
$r\simeq 13~R_\oplus$ which eventually accelerates to
$\sim 23~\km~\s^{-1}$ when leaving the domain of our
simulation. This outflow disperses the planet atmosphere at
a mass-loss rate of
$\dot{M} \simeq 5.5\times 10^{-10}~M_\oplus~\yr^{-1}$. Since
we assumed a constant high energy radiation output from the
host star, the mass-loss rates and the
hydrodynamic/thermodynamic profiles remain nearly constant
after reaching the quasi-steady state. We also note that we
did not put in any stellar wind from the host star, as the
current wind-less model fits the data reasonably well and is
preferred by Occam's Razor. However in a companion paper on
WASP-107 we will show that stellar winds may generate
Kelvin-Helmholtz instability that leads to fluctuations of a
photoevporative outflow.

Which mechanisms control the population of the \ext{He}
state? The bottom row in Figure~\ref{fig:wasp69b-profile}
compares the rates of different (de-)population processes
along the two particular streamlines (thickened curves in
Figure~\ref{fig:wasp69b-slice}). We compute the rate of
ionization, recombination, spontaneous decay, collisional
excitation and de-excitation as well as an advection
attenuation term $|\mathbf{v} \cdot \nabla
n(\ext{He})|$. Along the representative streamline presented
by the left column, the abundance of \ext{He} is determined
by the relatively stiff balance between the recombination
($\pos{He} + \e^-\rightarrow \ext{He}$) and the collisional
de-excitation at small radii $\lesssim 30~R_\oplus$. As
expected these two processes are efficient at higher
densities consistent with the law of mass
action. Photoionization of \ext{He} by soft FUV starts to
take over the destruction channel of \ext{He} where the
density of free electrons declines at higher altitudes. The
other channels have negligible importance: e.g. collisional
excitation from $1^1$S to the metastable state is more than
five orders of magnitude slower than recombination. On the
right column of Figure~\ref{fig:wasp69b-profile}, we show an
interesting streamline that crosses in the shadow of the
planet. The number density of \ext{He} soars in the shadow
because the photoionization of \ext{He} by soft FUV vanishes
here.

Beneath the base of the photoevaporative outflow
($r\lesssim 13~R_\oplus$), the temperature gradient between
the day-side and the night-side generates a slow "zonal"
circulation ($\sim 0.1~\km~\s^{-1}$). However, this region
has little observational effect on the overall \ext{He*}
observables which are mostly controlled by the much more
extended low density regions of the outflowing
atmosphere. We will return to this point shortly. Moving to
higher altitude, this day-night advection continues,
amounting to a $2-3\km~\s^{-1}$ blueshift at about
$20-40~R_\oplus$. Going further away from the planet, the
Coriolis effect starts to shape the outflow streamlines into
spiral curves resulting in blueshifts on the leading edge
and redshifts on the trailing edge. Considering that the
outflow is still primarily radial, increments in the
latitudinal velocity $|\Delta v_\theta|$ after traveling
through a radial distance $\Delta r$ can be estimated by,
\begin{equation}
  \label{eq:dv-coriolis}
  \begin{split}
    \Delta v_\theta \sim & ~ 2 \cos \phi\ \Omega_\K
    \int \d t\ v_r \simeq 2 \cos\phi\
    \Omega_\K \Delta r
    \\
    \simeq  & ~ 2.3~\km~\s^{-1}\ \cos\phi\
    \left(\dfrac{M_*}{M_\odot}\right)^{1/2} \\
    & \times
    \left(\dfrac{a}{0.05~\au}\right)^{-3/2}
    \left(\dfrac{\Delta r}{10~R_\oplus}\right)\ .
  \end{split}
\end{equation}
This estimation is confirmed by the velocity profile in the
top panels of Figure~\ref{fig:wasp69b-profile}: if we
compare the values at $r \simeq 40~R_\oplus$ and
$r\simeq 100~R_\oplus$, the difference in $v_\los$
(approximately equals to $v_\theta$ for this streamline) is
$\Delta v_\los \sim 11~\km~\s^{-1}$, and
eq.~\eqref{eq:dv-coriolis} yields $\sim
10.8~\km~\s^{-1}$. This effect re-distributes \ext{He} atoms
in the velocity space and broadens the observed \ext{He}
line profiles as we will see shortly.

\subsection{Comparison with Observations}

Figure \ref{fig:wasp69b-spec} shows our synthetic observations
of both the line profiles and light curves of WASP-69b
\citep{2018Sci...362.1388N,2020AJ....159..278V}. We have
binned the light curve data from \citet{2020AJ....159..278V}
for better clarity and the uncertainty represents the
standard deviation within each phase bin. Our fiducial model
of WASP-69 seems to fit both the spectroscopic and
photometric observations well simultaneously. In particular,
the synthetic line profile reproduced the subtle blueshift
of the peak absorption, the overall line depth, and the
relative ratio between the lines of this
triplet. Numerically, \citet{2018Sci...362.1388N} reported a
net blueshift of $-3.58 \pm 0.23~\km~\s^{-1}$. This
blueshift was based on fitting Gaussians to the observed
line profiles; however, as we argued in the previous
section, kinematic shift of the outflow introduces
significant distortion of the spectral shape. Instead of
fitting Gaussian to our line profile, we compared our
simulations directly to the line profile itself which shows
great agreement. We also use a different way to measure the
blueshift: we report the blueshift of the peak absorption
relative to a line-ratio-weighted average of the rest-frame
line center for the two longer wavelength transistions that
are usually blended together.

\citet{2018Sci...362.1388N} hinted at the possibility of a
comet-like tail trailing behind WASP-69b. The basis of their
suggestion is that additional \ext{He} absorption can still
be seen $\sim 20~\min$ after the nominal egress of the
planet. The higher precision, better temporally sampled
photometric data from \citet{2020AJ....159..278V} nontheless
favors a symmetric transit. The symmetric transit shape
(lower panel of Figure~\ref{fig:wasp69b-spec}) does not
support an extended comet-like tail. Our simulation seems to
be more consistent with \citet{2020AJ....159..278V}, the
photoevaporative outflow of WASP-69b in our fiducial model
is largely symmetric between the leading and trailing edge,
hence it produces a more symmetric transit shape. We note
that an extended comet-like tail will also introduce
significant distortion to \ext{He} line profile (see our
Companion paper on WASP-107b for strong comet-like tail
generated by strong stellar wind in that system). Here in
the case of WASP-69b, our fiducial model produces a good fit
the resolved line profile \citep{2018Sci...362.1388N} while
it does not need to invoke a prominent comet-like tail.

Another point we would like to emphasize is that a
significant part of the \ext{He} absorption for WASP-69
seems to be produced by an extended, optically thin
($\tau < 1$) outer layer of the photoevaporative outflow. In
Figure~\ref{fig:wasp69b-ext}, we show the mid-transit
extinction ($1-\exp^{-\tau}$) at three different wavelengths
near the \ext{He} transitions. The outer regions (10s of
$R_\oplus$) contribute significantly to the overall
extinction thanks to their extended area and the slow
decrease of \ext{He} number density in the outflow. Because
of the unsaturated optical depth, the line ratios between
the \ext{He} triplet are close to $1:3:5$ i.e. their quantum
degeneracies. More accurately, the line ratios are close to
$1:8$ as the longer two lines are blended by kinematics and
thermal broadening. This suggests that the line ratios
between the \ext{He} triplet can be a diagnostic of the
number density in the outflow. If most of the \ext{He}
absorption is due to higher-density region where one of the
line may saturate first, the line ratio will deviate from
the quantum degeneracy ratio; this would tell us about the
density of the outflow region in a model-independent way
\citep[see also discussions in][]{2018A&A...620A..97S}. This does not seem
to be the case for WASP-69b, as most \ext{He} absorption
happens in lower density regions.

It is worth noting that, due to heavy computational cost of
our 3D simulations, we could not afford to numerically fit
the data with multiple simulation runs. Instead, our
fiducial model serves as a validation our self-consistent 3D
hydrodynamic simulations: with reasonable assumptions of the
planetary/stellar properties and high energy SED, we can at
least qualitatively reproduce the various \ext{He}
observables. That said, the degree to which our simulation
agrees with observations is quite encouraging if not
remarkable. After this validation of model, we stand at a
position to perturb our fiducial model and investigate how
the \ext{He} observables are impacted by various factors
that control the underlying photoevaporative outflow and
\ext{He} population in the following section.

\section{Parametric Study}
\label{sec:parameter-study}

How do the photoevaporative outflow and the resultant
\ext{He} observables depend on key parameters in our
simulations? In this section, we explore the impact of
simulation dimensionality, XUV flux levels, host star
spectral type, and planet surface gravity. This is done by
perturbing validated fiducial model in these
parameters. Before that, we did a further convergence
test. We reran the fiducial model with a much finer
simulation grid of
$N_{\log r}\times N_\theta\times N_\phi = 192\times
192\times 128$ (versus the fiducial model,
$N_{\log r}\times N_\theta\times N_\phi = 144\times
128\times 64$). The resultant \ext{He} observables in the
convergence test are almost identical
(Figure~\ref{fig:wasp69b-spec}) to the much faster fiducial
model. This gives us confidence that the adopted spatial
grid is fine enough to resolve the photoevaporative outflow
on WASP-69b.

\subsection{Dimensionality}

To test how our model depends on the spatial dimensions of
the simulations, we ran a 2D model with axisymmetry (about
the $\phi$ axis i.e. $N_\phi = 1$) while keep all system
parameters the same as the fiducial model.  The Coriolis
forces is not captured in this 2D simulation while stellar
gravity and orbital centrifugal forces are still involved. A
reference 1D spherical symmetric model is also implemented
using the $\theta = \pi / 2, \phi = \pi / 2$ radial line and
removing the $\theta$ and $\phi$ components of the velocity.

Figure~\ref{fig:wasp69b-spec} compares the synthesized line
profiles and light curves with all three dimensionality
models.  The 1D spherically symmetric model suffers from the
loss of all non-radial kinematic information. It is clearly
inconsistent with the observed line profile with no
blueshift and limited kinematic broadening. The 2D
axisymmetric test is able to capture the day-night
advection. It shows good agreement with the observed line
profile. The 3D fiducial model further modifies the line
profile by including the Coriolis force. In this case of
WASP-69b such a modification is quite subtle, which again
testifies that the photoevaporative outflow on WASP-69b is
largely symmetric between the leading and trailing
edge. Again see our companion paper on WASP-107b for how
this symmetry is broken by the inclusion of stellar winds.

The three models from 1D through 3D have almost identical
equivalent width ($\mean{W_\lambda}\sim 3.1~\ang$) and light
curves. This stresses the importance of spectrally resolving
the \ext{He} line profiles which are seen to vary the most
between dimensions. The mass-loss rate are again quite
similar between 3D and 2D models at about
$\dot{M} \simeq 5.5\times 10^{-10}~M_\oplus~\yr^{-1}$. The
mass-loss rate in our 1D model is off
($\dot{M} \simeq 6.9 \times 10^{-10}~M_\oplus~\yr^{-1}$)
because it assumes perfect spherical symmetry. However, the
streamlines in Figure \ref{fig:wasp69b-slice} are clearly
non-radial. We also compare our results with that from a 1D
isothermal model
\citep{2018ApJ...855L..11O,2020AJ....159..278V} of

$9.5 \times 10^{-10}~M_\oplus~\yr^{-1}$
($\simeq 3\times 10^{-3} ~M_{\rm Jup}~{\rm Gyr}^{-1}$) at an
assumed temperature of 12000K. The results are in rough
agreement with differences arising from more careful
treatment of the hydrodynamics, thermodynamics and radiative
transfer.

\subsection{XUV Flux Intensity}
\label{sec:XUV}

\begin{deluxetable}{lccccc}
  \tablecolumns{6} 
  \tabletypesize{\scriptsize}
  \tablewidth{0pt}
  \tablecaption{Radiation fluxes for different model host
    stars based on the fiducial model of WASP-69b 
    \label{table:wasp69b-var-star-flux}
  }
 \tablehead{
    \colhead{Star} \vspace{-0.25cm} &
    \colhead{$F_{20}$} &
    \colhead{$F_{15}$} &
    \colhead{$F_{13}$} &
    \colhead{$F_{13}$} &
    \colhead{$F_{13}$}
    \\
    \colhead{type} &
    \colhead{$(2~\eV)$} &
    \colhead{$(7~\eV)$} &
    \colhead{$(12~\eV)$} &
    \colhead{$(20~\eV)$} &
    \colhead{$(40~\eV)$} 
  }
  \startdata
  F & 2.8 & $1.6\times 10^4$ & 4.7 & 4.0 & 2.0 \\
  G & 1.4 & $8\times 10^2$ & 2.4 & 2.0 & 2.5 \\
  M & 0.11 & 0.6 & 0.1 & 0.12 & 0.36 \\
  \enddata
   \tablecomments{For simplicity, $F_{N}(h\nu)\equiv
   F(h\nu) / (10^{N}~{\rm cm}^{-2}{\rm s}^{-1})$,
   calibrated for the value at the planet orbit without
   any extinction. }
\end{deluxetable}

\renewcommand{\arraystretch}{1.2}
\begin{deluxetable*}{lccccc}
  \tablecolumns{6} 
  \tabletypesize{\scriptsize}
  \tablewidth{0pt}
  \tablecaption{Results of various models, based on 
    the fiducial model for WASP-69b 
    \label{table:wasp69b-var}
  }
 \tablehead{
    \colhead{Model} \vspace{-0.25cm} &
    \colhead{Description} &
    \colhead{$\dot{M}$} &
    \colhead{$\mean{W_\lambda}$} &
    \colhead{$\Delta v_{\rm peak}^\dagger$} &
    \colhead{FWHM$^*$}    
    \\
    \colhead{  } &
    \colhead{} &
    \colhead{($10^{-9}~M_\oplus~\yr^{-1}$)} &
    \colhead{($10^{-2}~\ang$)} &
    \colhead{($\km~\s^{-1}$)} &
    \colhead{($\km~\s^{-1}$)} 
  }
   \startdata
   69-0 & 3D Fiducial &$0.55$ & $3.16$ & $-2.1$ & $17.3$ \\ 
   69-0-2D & 2D Test (fiducial parameters) &$0.56$ 
   & $3.10$ & $-2.0$ & $16.6$ \\ 
   69-0-1D & 1D Test (fiducial parameters) &$0.69$ 
   & $3.14$ & $0.07$ & $16.2$ \\
   \hline
   69-1 & $10\times $ Flux for soft FUV ($h\nu = 7~\eV$)
   & $0.61$ & $1.75$ & $-1.7$ & $14.0$ \\
   69-2 & $10\times $ Flux for LW ($h\nu = 12~\eV$)
   & $0.56$ & $3.18$ & $-2.1$ & $17.4$ \\    
   69-3 & $10\times $ Flux for soft EUV ($h\nu = 20~\eV$)
   & $0.66$ & $4.13$ & $-2.4$ & $24.2$ \\ 
   69-4 & $10\times $ Flux for hard EUV ($h\nu = 40~\eV$)
   & $2.50$ & $5.62$ & $-3.8$ & $48.1$ \\
   69-5 & $10\times $ Flux for X-ray ($h\nu = 3~\keV$)
   & $0.55$ & $3.12$ & $-2.0$ & $17.6$ \\
   \hline
   69-6 & $1/2$ Planet mass ($M_\p = 41.3~M_\oplus$)
   & $0.93$ & $4.25$ & $-2.1$ & $17.2$ \\
   \hline
   69-F & Fiducial Model with F-type host & $0.70$ & $0.04$
   & $-0.8$ & $9.0$ \\ 
   69-G & Fiducial Model with G-type host &  $0.64$ & $0.51$
   & $-1.5$ & $10.3$ \\    
   69-M & Fiducial Model with M-type host & $0.05$ & $1.40$
   & $-0.6$ & $12.9$ \\
   \enddata
   \tablecomments{The values are time averages taken over
     the last $15~\tau_\dyn$ of the simulations.
     Fluctuations are negligible for all models in the
     table. \\
     $\dagger$: Shifts (postive values for redshifts and
     vice versa) of the right-hand-side peaks in the
     velocity space, compared to the line-ratio-averaged
     line center of $\lambda = 10833.29~\ang$. 
     \\
     $*$: Full-width half-maximum of the longer-wavelength
     peak in the velocity space.
     \\
   }
 \end{deluxetable*}
 
\begin{figure}
  \centering
  \includegraphics[width=3.4in, keepaspectratio]
  {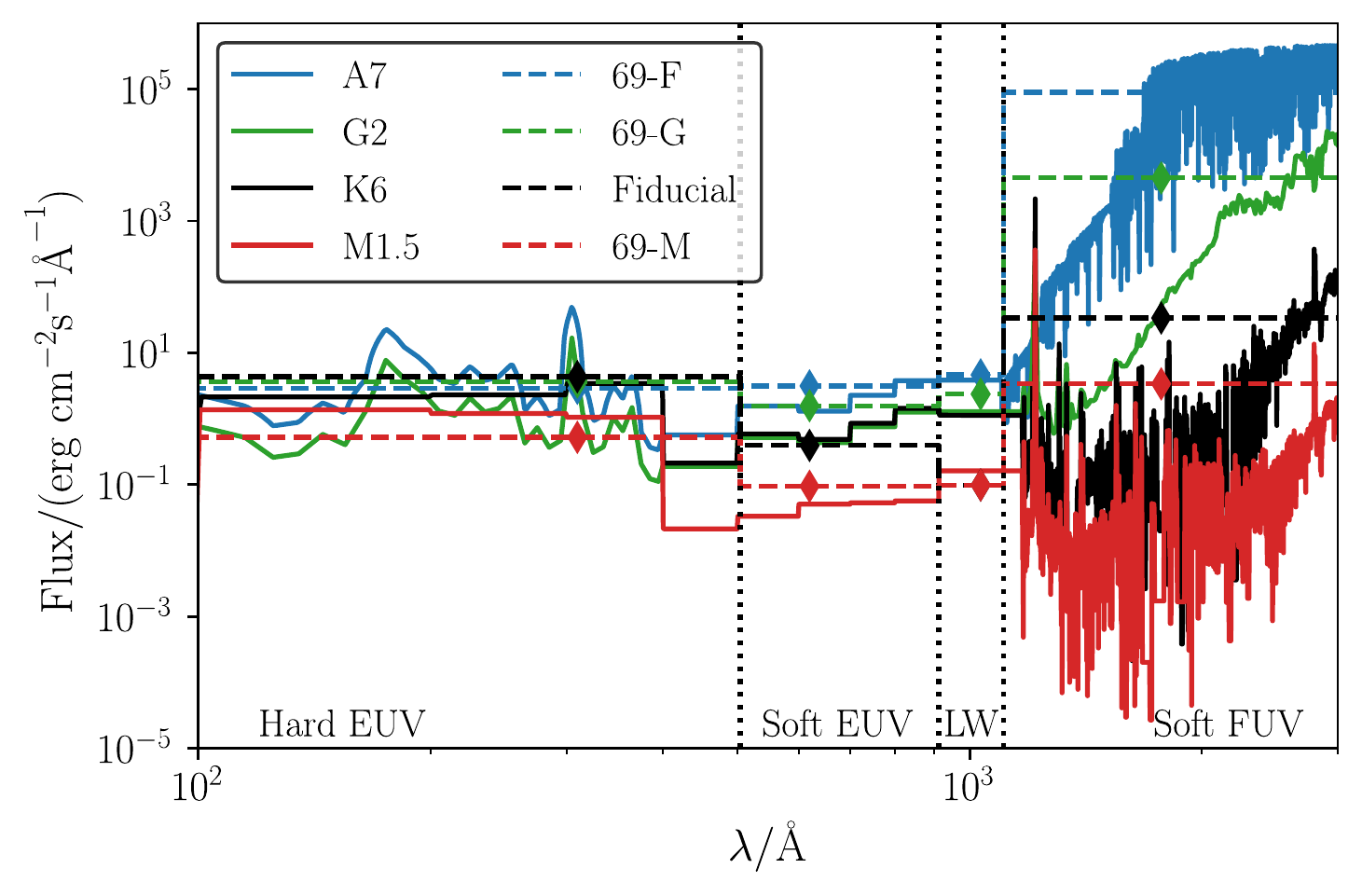}
  \caption{Spectral energy distributions of the incidenting
    high-energy photon fluxes, for the models with Type F,
    G, K and M host stars (Models 69-F, 69-G, fiducial model
    69-0, and 69-M), presented in dashed lines. Diamonds
    overplotted on the dashed lines mark the exact energy of
    incident photons in the simulations (as represetatives
    of the corresponding energy bands; see
    \S\ref{sec:method-chem}). For reference, the model
    spectra of stellar spectral types A7, G2, K6 and M1.5
    (based on the compilation in
    \citealt{2019ApJ...881..133O}) are shown in solid
    curves. Vertical dotted lines mark the boundary between
    different energy bands: ``Hard EUV'' for
    $h\nu > 24.6~\eV$ photons that can ionize helium; ``Soft
    EUV'' for $13.6<(h\nu/\eV)<24.6$ photons that can ionize
    hydrogen ; ``LW'' (short for Lyman-Werner) for
    $11.2<(h\nu/\eV)<13.6$ photons that can photodissociate
    \chem{H_2}; ``Soft FUV'' for $h\nu < 7~\eV$ photons. }
  \label{fig:wasp69b-spec-mod}
\end{figure}

\begin{figure}
  \centering
  \includegraphics[width=3.4in, keepaspectratio]
  {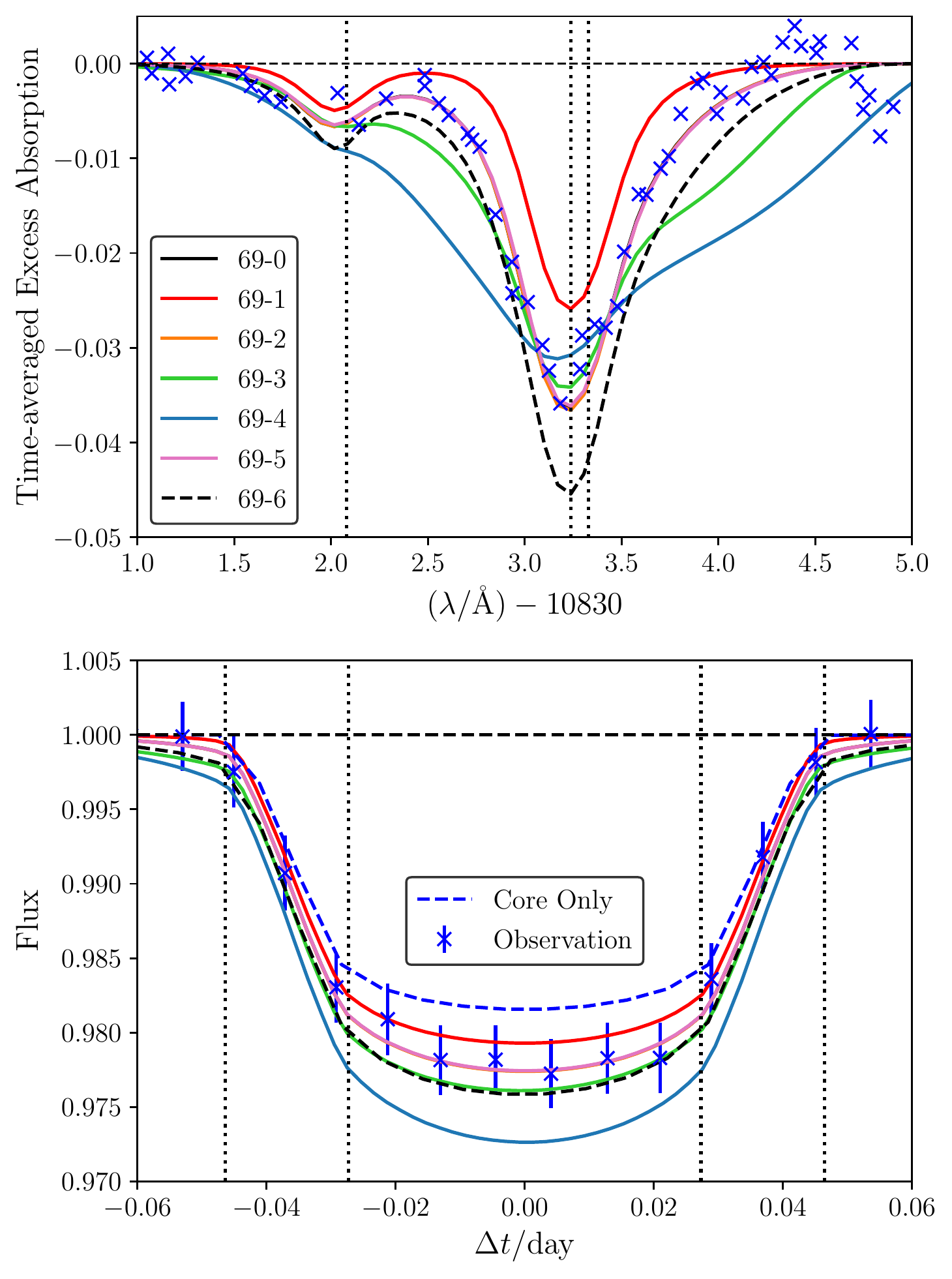}
  \caption{Similar to Figure~\ref{fig:wasp69b-spec} but for
    the models 69-0 through 69-6 in
    Table~\ref{table:wasp69b-var}. The fiducial model 69-0
    is also included for reference. }
  \label{fig:wasp69b-var-spec}
\end{figure}

\begin{figure}
  \centering
  \includegraphics[width=3.4in, keepaspectratio]
  {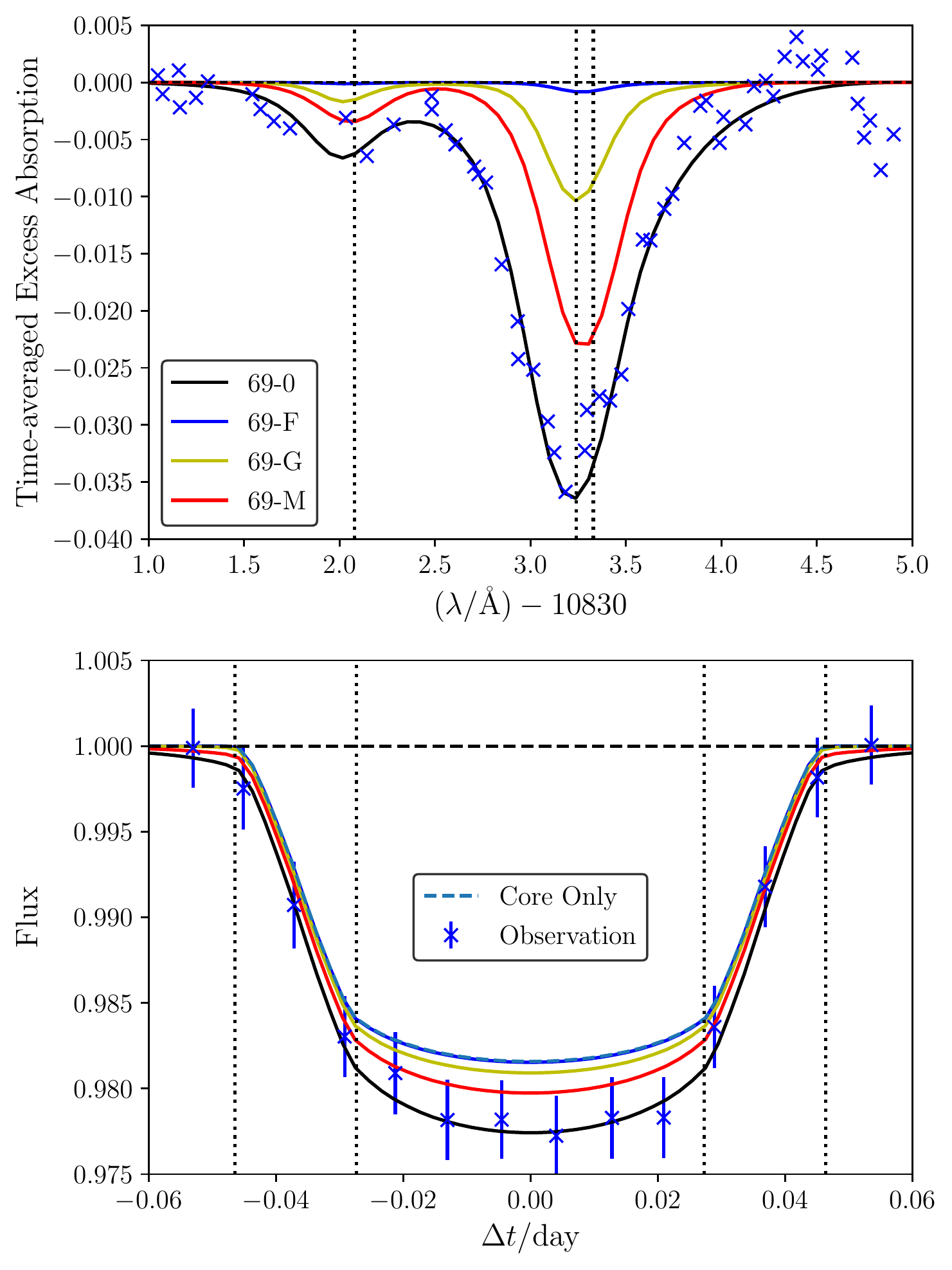}
  \caption{Similar to Figures~\ref{fig:wasp69b-spec} and
    \ref{fig:wasp69b-var-spec}, but presenting the models
    69-F, 69-G, and 69-M for different host star spectral
    types in Table~\ref{table:wasp69b-var}.  Note fiducial
    model 69-0 is a K star.}
  \label{fig:wasp69b-var-star-spec}
\end{figure}

\begin{figure}
  \centering
  \includegraphics[width=3.4in, keepaspectratio]
  {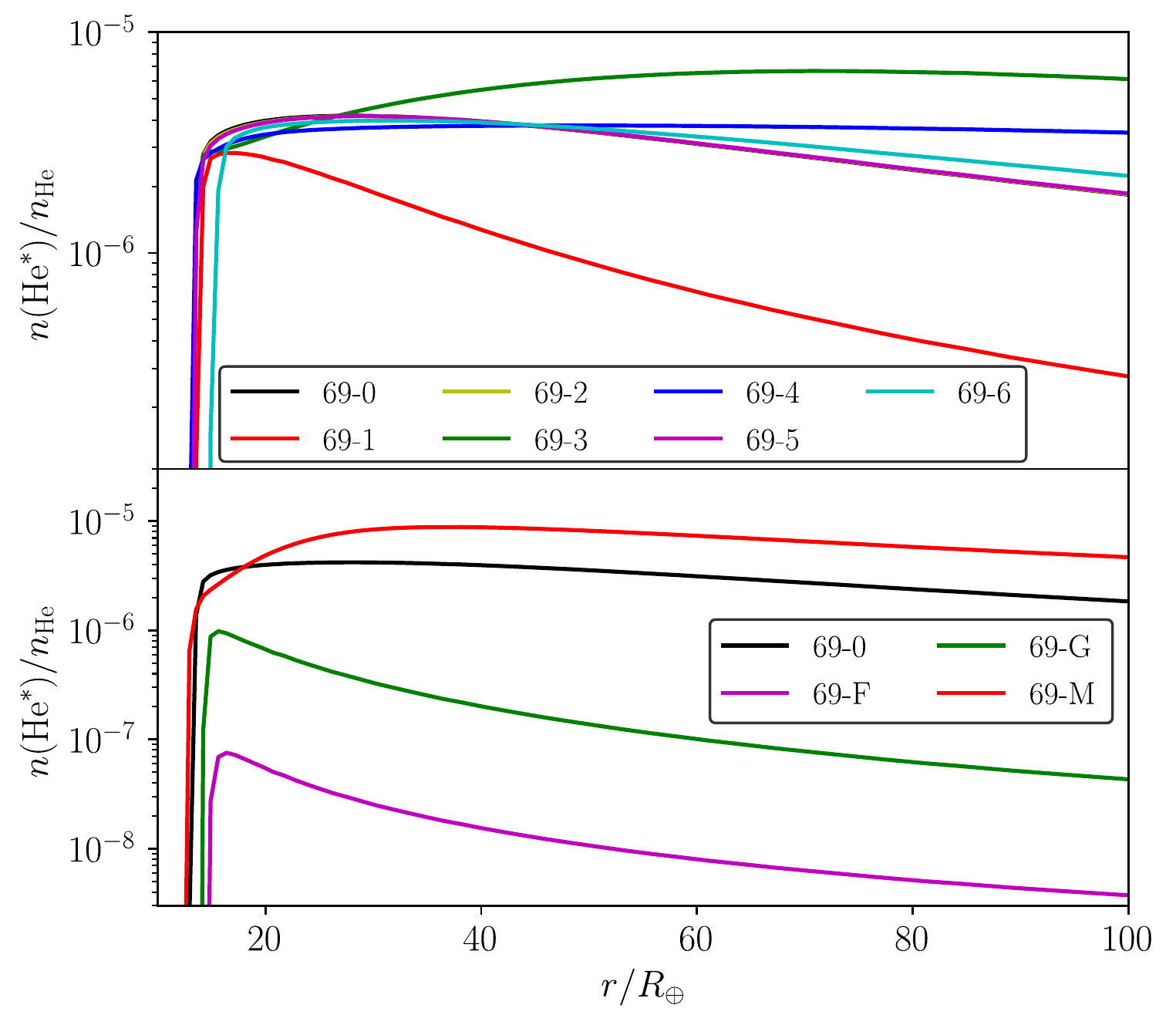}
  \caption{Relative abundance profiles of \ext{He}
    (relateive to total helium nucleus density
    $n_{\rm He}$), measured along the radial line with
    $\theta = \pi/4$, $\phi = 0$ in each simulation domain,
    for models described in
    Table~\ref{table:wasp69b-var}. {\bf The upper panel}
    shows models 69-0 and 69-1 through 69-6, while {\bf the
      lower panel} specifically compares the results of
    different types of host stars (Model 69-0 for the K star
    WASP-69, and Models 69-F, 69-G and 69-M for F, G, M
    stars respectively). }
  \label{fig:wasp69b-var-profile}
\end{figure}

Photoevaporative outflows are driven by high energy
radiation from the host star. Moreover, the population of
\ext{He} states are also controlled by the critical balance
high energy photons of different energy bins. We examine the
impact of high energy radiation in each energy bin by
perturbing the fidual model. The amount of high energy
radiation a star outputs is variable depending on the
evolution stage, activity and spectral types of the host
star. Direct measurements are also lacking as the XUV
measurements have to be performed in space. Therefore, in
our Models 69-1 to 69-5 (soft FUV for 69-1, LW for 69-2,
soft EUV for 69-3, hard EUV for 69-4, X-ray for 69-5), we
bump up the flux level in each high energy bin by a whole
order of magnitude to reflect the intrinsic variation in
high energy flux level. We summarize key \ext{He}
observables in Table~\ref{table:wasp69b-var}; we also show
the synthetic line profiles and light curves in
Figure~\ref{fig:wasp69b-var-spec} and the relative abundance
of \ext{He} as a function of radius in
Figure \ref{fig:wasp69b-var-profile}.

The \ext{He} line profiles are controlled mainly by the FUV
(adverse effect) and the EUV bands (positive effect); the LW
and X-ray bands only play minor roles under the ``typical''
host star conditions. More specifically, the relative
abundance of \ext{He} is suppressed by soft FUV because it
photoionizes the \ext{He} states, thus significantly
reducing its population and the \ext{He} absorption depth.
In Model 69-1 ($\times10$ soft FUV flux), the over-all
mass-loss rate is enhanced by a few percent thanks to extra
energy deposited into the atmosphere. However the stronger
soft FUV flux significantly lowered the number density of
\ext{He} at a larger radial extent ($r\gtrsim
20~R_\oplus$). The \ext{He} line profile depth and the light
curve depth are both reduced by about a factor of two
($3.16~\ang$ to $1.75~\ang$). The line width also decreased
(FWHM from $17.3~\ang$ down to $14.0~\ang$) because the
high-altitude region with a higher velocity dispersion
contribute less to the \ext{He} extinction.

In Model 69-2 ($\times10$ LW flux) the \ext{He} observables
are mostly unaltered from the fiducial model. This is
because the LW band is intrinsically narrow thus only amount
to a very small fraction of the overall high energy
radiation flux. Moreover, most molecular H$_2$ are already
dissociated at the $\sim 10^4$K in our simulations.

At higher EUV fluxes (Model 69-3 and 69-4), the much faster
outflows not only bring more \ext{He} into the exosphere but
also spread them out in velocity space effectively
broadening the line profiles. This confirms our earlier
picture that EUV flux are most effective in driving
photoevaporative outflows (WD18).

Finally, X-ray seems to play a secondary role in
photoevaporation and \ext{He} observables. Model 69-5
($\times10$ X-ray flux) has very similar observables as the
fiducial model. Although X-rays photons are very energetic,
they also have much smaller cross sections than EUV
photons. As a result, X-ray penetrate deeper into the
atmosphere where collisional processes and dust particles
quickly convert the X-ray energies to infrared radiation
that escapes easily. This limits the heating potential of
X-ray. We reached a similar conclusion in WD18.

\subsection{Host Spectral Type}
\label{sec:parameter-host}

Among the handful of reported \ext{He} detections, 4 out of
6 are planets around K-type hosts
\citep{2018Natur.557...68S, 2018A&A...611A..36V,
  2018Sci...362.1384A, 2018A&A...620A..97S,
  2020ApJ...894...97N,
  2019A&A...629A.110A}. \citet{2019ApJ...881..133O} 
confirmed that K-stars, at least in 1D isothermal models,
may be at the sweet spot of FUV/EUV flux balance that best
promote the \ext{He} population and thus observablity. This
section re-evaluates such a claim with our 3D hydrodynamic
simulation coupled with self-consistent thermodynamics and
radiative transfer.

We set up three additional models 69-F, 69-G, and 69-M,
whose luminosities in different high energy bins emulate
typical F-type, G-type, and M-type main-sequence stars based
on the compilation of \citealt{2019ApJ...881..133O}. We
remind the reader the fiducial model 69-0 has a K-star
SED. The flux in each high energy bin is summarized in
Table~\ref{table:wasp69b-var-star-flux}). Broadly speaking,
F-type and G-type stars output similar levels of soft and
hard EUV fluxes as K-type stars, however their FUV
luminosities are significantly higher by a factor of
$\sim 3000$ and $\sim 130$ respectively. For a typical M
star, fluxes in all high energy bands are lower by about one
order of magnitude.

\ext{He} observables of these models are summarized in
Table~\ref{table:wasp69b-var} for their mass-loss rates and
a few key diagnostics etc. We show the line profiles and
light curves in Figure~\ref{fig:wasp69b-var-spec} and the
radial distribution of \ext{He} in
Figure \ref{fig:wasp69b-var-profile}. With a soft FUV flux
$\sim 3000$ times higher than our fiducial model, F-type
stars significantly suppress the population and
observability of \ext{He} with an equivalent width reduced
by almost two orders of magnitude (3.16 to 0.04
\ang). However, the mass loss rate of the photoevaporative
outflow is similar to that of the fiducial model. Again,
photoevaporation is driven mostly by EUV which have similar
flux levels between F and K stars. On the other hand, for a
typical M star host, whose higher energy flux levels are
weaker in all bands, the mass loss rate is reduced by one
order of magnitude. Nonetheless, the equivalent width of
\ext{He} in the transmission spectrum only decline by a
factor of 2 (3.16 to 1.4 \ang). Again this is because its
much weaker soft FUV flux allows proportionally more
\ext{He} to exist in the outflow
(Figure \ref{fig:wasp69b-var-profile}). G-star represents some
middle ground, its $\sim 130$ times higher soft FUV flux
suppress the equivalent width \ext{He} by a factor of 6.

In summary, our findings suggest that K-star planet hosts
are indeed favorable targets for \ext{He} observations
consistent with the suggestion of
\citet{2019ApJ...881..133O}. The high-energy SED,
nonetheless, is expected to change significantly as a
function of host star age and activity level. The
suppression factor of \ext{He} around G and M type stars are
often only a factor of a few. We encourage observers to keep
them in their target list particularly the young and active
ones. We also predict that there will be more reports of
\ext{He} detection around G and M type hosts soon. Another
important point we would like to stress is that the depth of
\ext{He} line profile {\bf can not} be translated to the
underlying mass loss rate without knowing the high energy
SED of the host star. In other words, measuring the XUV SED
of the host star directly is crucial for correctly
interpreting the \ext{He} observations.

\subsection{Surface gravity}
\label{sec:parameter-grav}

The mass-loss rate of photoevaporation depends quite strong
on the depth of gravitational potential well of the
planet. A shallower potential allows faster outflow with the
same high energy irradiation. In Model 69-6 we adjusted the
planet interior such that the planet mass is reduced by half
while keeping the transit radius the same. This effectively
lowers the surface gravity of the planet by a factor of
two. The mass loss rate in Model 69-6 increases to
$\dot{M}\simeq 9.3\times 10^{-10}~M_\oplus~\yr^{-1}$ which
is $\sim 70~\%$ larger than the fiducial model. This larger
mass loss rate can be decomposed into an increase in the
terminal outflow velocity by $\sim 10~\%$ and an increase of
the outflow density by $\sim 60~\%$.

The \ext{He} line profile depth responds to this increase of
mass loss rate sub-linearly. In
Figure~\ref{fig:wasp69b-var-spec}, the line profile has
$\sim 40~\%$ larger depth than that in the fiducial model,
while the equivalent width increases by about $\sim
30~\%$. However, the \ext{He} line profile maintains a
similar morphology with the peak velocity shift and the FWHM
unchanged from the fiducial model
(Table~\ref{table:wasp69b-var}). In short, puffier, low
surface gravity planets are more likely to undergo strong
photoevaporative mass loss and should prove great target for
\ext{He} observations.
 
\section{Variability and Stellar Flares}
\label{sec:parameter-flares}

\begin{figure}
  \centering
  \includegraphics[width=3.5in, keepaspectratio]
  {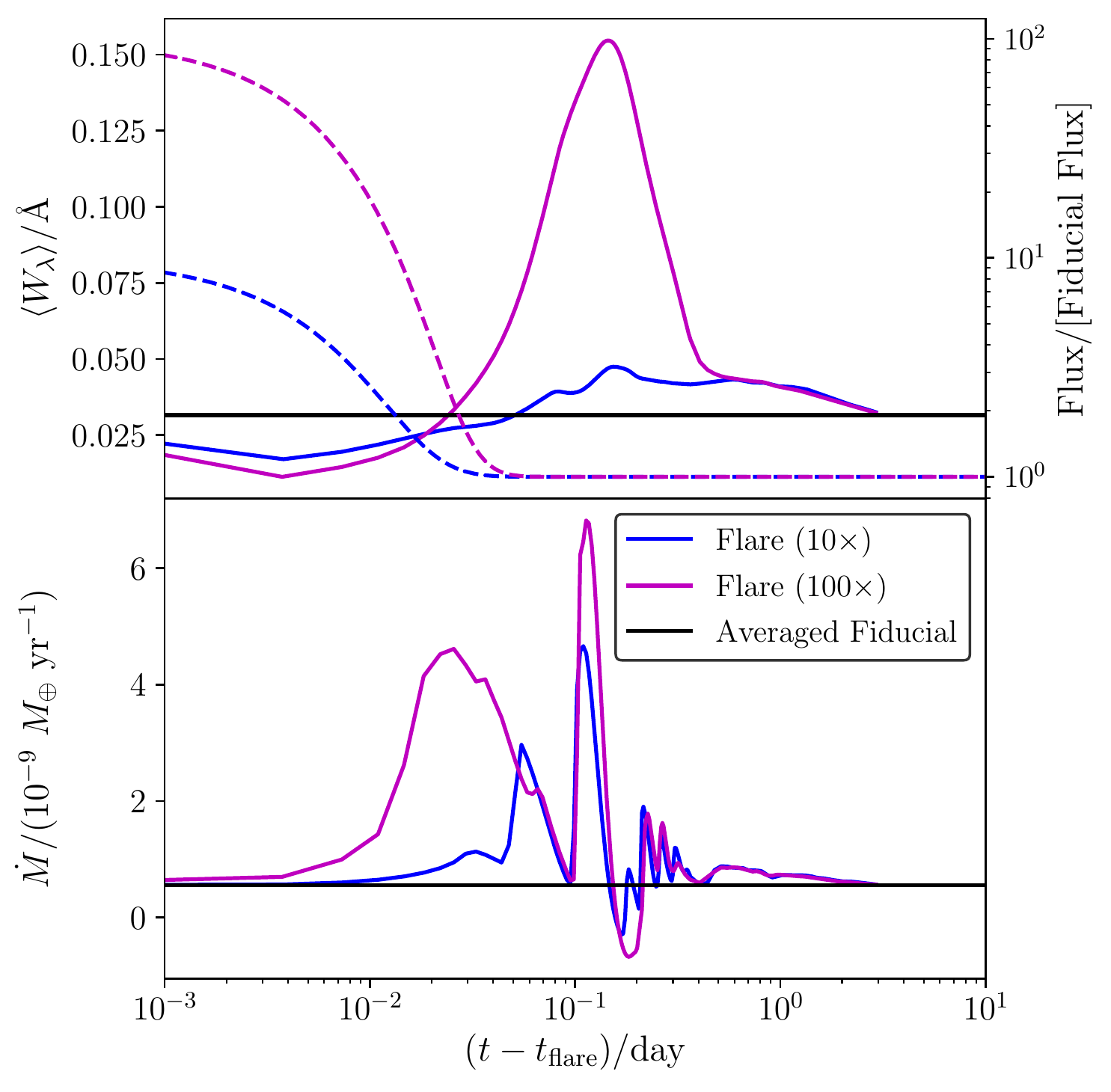}
  \caption{Temporal variation of the two models (based on
    Model 69-0) with flares that increase all high-energy
    radiation by $10$ and $100$ times uniformly
    (\S\ref{sec:parameter-flares}). {\bf The top panel}
    shows the equivalent widths of the metastable helium
    absorption by solid curves. The two dashed lines, which
    should read the right ordinate, indicate the evolution
    of radiation fluxes in all energy bands.  The mass-loss
    rates are shown in {\bf the lower panel}. Time-averaged
    values for the fiducial model are also plotted for
    reference. }
  \label{fig:wasp69b-flare}
\end{figure}

\begin{figure}
  \centering
  \includegraphics[width=3.5in, keepaspectratio]
  {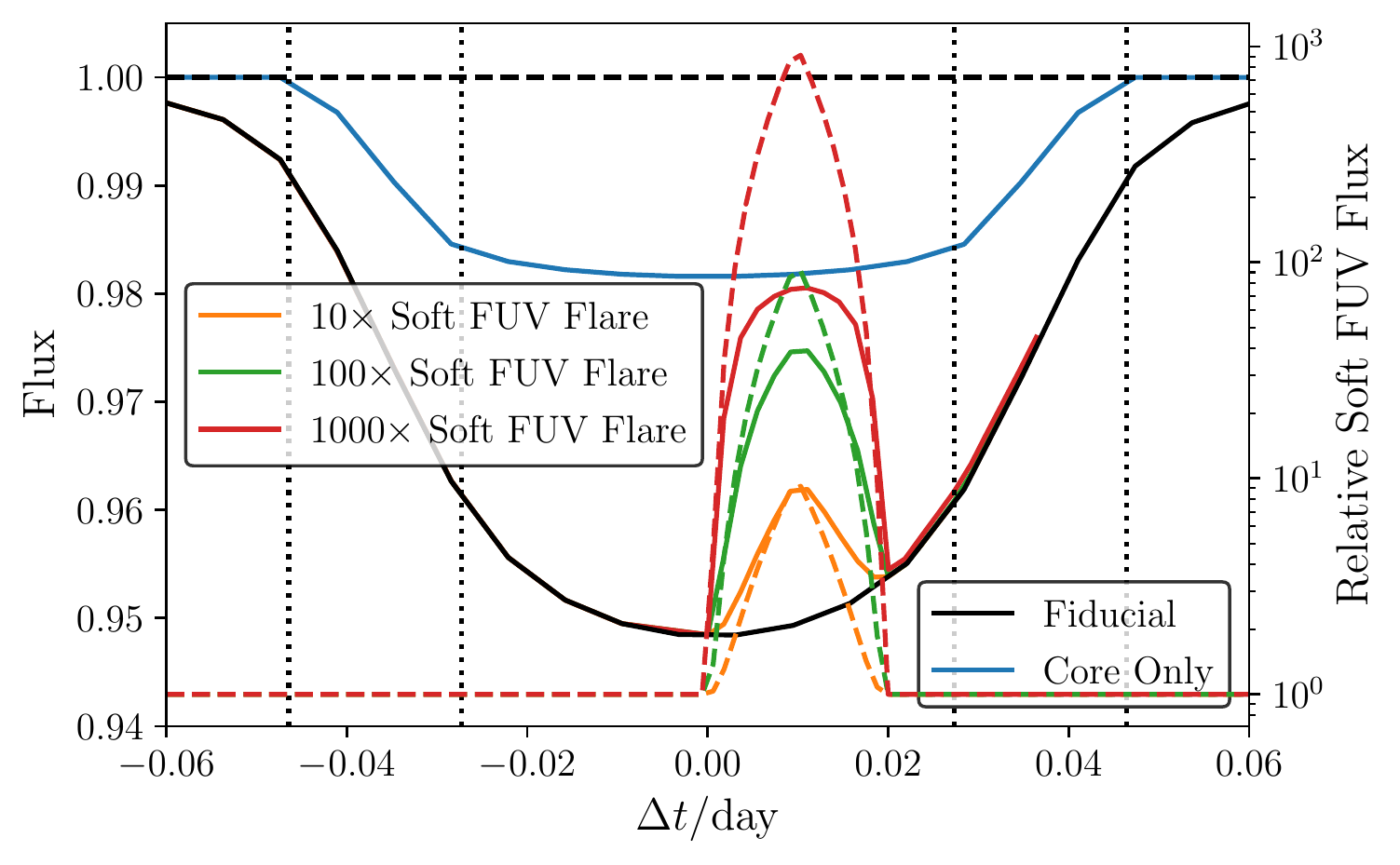}
  \caption{Slimilar to the lower panel of
    Figure~\ref{fig:wasp69b-spec}, showing the light curves
    following {\bf soft-FUV-only} flares that start at
    mid-transit ($\Delta t = 0$) and terminating after
    $0.5~\hr$. Different flare intensities are indicated by
    colors. The light curves in solid lines should read the
    left ordinate, while the dashed lines are the flare
    shapes and should be read with the right
    ordinate. Soft-FUV-only destroys \ext{He} while does not
    significantly perturb the photoevaporative outflow, thus
    only produces a decrease in \ext{He} line depth that is
    seen in WASP-69b. Nontheless, we do not believe this is
    the explanation for the observed variability (see
    \S\ref{sec:parameter-flares} for detail).}
  \label{fig:wasp69b-flare-lc}
\end{figure}

During one of the transit of WASP-69b in
\citet{2018Sci...362.1388N}, the \ext{He} line profile
experienced a $\sim 30\%$ drop in magnitude that lasted for
about 20 minutes. This variability could well be
instrumental in origin, but here we explore an alternative
explanation that it is generated by a stellar flare on the
host star.

Solar/stellar flares are associated with the surface
magnetic activity of the host star. Their amplitudes can
range below a percent to even orders of magnitude in extreme
cases \citep[``superflares''][and references
therein]{2020AJ....159...60G}.  The sudden rise of
luminosity is often followed by exponential decays to
nominal flux level on minutes or hours timescale. To
investigate the consequences of flaring events on \ext{He}
observables, we first inject a simple flare model in which
fluxes across {\it all} energy bands energy bands increase
by a factor 10 and 100 times which then quickly decay to the
quiescent state exponentially with a timescale of $500~\s$.

The temporal response of our fiducial model to these flares
are shown in Figure \ref{fig:wasp69b-flare}. Independent of
the energy injected, the common mode of response is as
follows. Before the dynamics of the outflowing atmosphere
can fully respond to the flares, the photoionization of
\ext{He} by soft FUV photons reduces the number density of
\ext{He} and cause a significant decrease in the equivalent
width. It is only after the dynamical timescale
$t - t_{\rm flare}\sim \tau_\dyn \sim$ hours that the
flare-generated surge of photoevaporative mass-loss reach
the higher altitudes where most of \ext{He} absorption
occurs. As a result, the equivalent width of \ext{He}
increases after the first hour or so and remains high for
several hours. Looking at the mass-loss rate across the
outer boundary of the simulation (lower panel of Figure
\ref{fig:wasp69b-flare}), it experiences several
oscillations on dynamical timescales as the systems response
to the increased flux from the flare. The equivalent width
(and other observables) however are the spatially integrated
quantities, thus it effectively smears out most of these
oscillations and has a much smoother variation (upper panel
of Figure \ref{fig:wasp69b-flare}). Comparing with
variability seen in WASP-69b \citep{2018Sci...362.1388N}, a
flare that simultaneously raises all high energy radiation
does not appear to be a good explanation. This is because it
should be observed as a decrease followed by an increase of
\ext{He} absorption rather than the decrease only in the
observations \citep{2018Sci...362.1388N}.

We therefore explore a different flare model in which {\it
  only} the soft FUV band. We do not have observational
support that flares of this kind exist. We explore this
rather contrived scenario just out of curiosity. Remember
from \S \ref{sec:XUV} that the soft FUV primarily suppresses
the \ext{He} abundance by photoionization without
significantly changing the overall kinematics. A
soft-FUV-only flare may produce the observed decrease of
\ext{He} line depths. We setup three extra simulation runs,
again based on the fiducial model of WASP-69b. We put in
soft FUV flares ($10$, $100$ and $1000$ times the nominal
value) that start at the middle of the transits and last for
$30~\min$.  In Figure~\ref{fig:wasp69b-flare-lc}, we can see
that the light curves respond to these soft-FUV-only flares
quickly. In order to reproduce the $\sim$30\% variation, a
soft FUV flare between $10-100\times$ the nominal level is
required. This should be readily observable in the
\ion{Ca}{ii} H, K lines of the CARMENES spectra in
\citet{2018Sci...362.1388N}. Nevertheless, enhanced activity
was not observed in the spectra during or preceding the
observed light curve variation (private communications,
Nortmann). Afterall, stellar flares do not seem to be a
viable explanation of the temporal variability seen the
\ext{He} line profiles of \citet{2018Sci...362.1388N};
instrumental effect is perhaps a better solution.

\section{Summary}
\label{sec:summary}

In this work, we simulate the ionized mass loss of close-in
exoplanets and the metastable helium absorption during the
planetary transit. We produce synthetic spectrally resolved
line profiles and the light curves in a narrow filter band
around the \ext{He} transitions. Dynamics of such synthesis
requires 3D hydrodynamic simulations of photoevaporating
planetary atmospheres; non-equilibrium thermochemistry and
ray-tracing radiative transfer are co-evolved with the
hydrodynamics. The processes that populate and depopulate
the metastable state of neutral helium are included in a
thermochemical network and solved efficiently on GPUs.

With reasonable assumptions about the system parameters and
high energy SED of WASP-69, we find a plausible model that
launches a photoevaporative outflow with a mass-loss rate of
$\dot{M}\simeq 5.5\times 10^{-10}~M_\oplus~\yr^{-1}$. The
model yields a spectrum and a light curve that are in
remarkable agreement with the observations in terms
equivalent width, line-ratios, blueshift and line broadening
\citep{2018Sci...362.1388N,2020AJ....159..278V}. Inside this
outflow, metastable helium is formed almost solely by
recombination. Its destruction is mainly due to collisional
de-excitation at small radii where the density is high and
photoionization by FUV photons at outer lower-density
regions.

With this fiducial model of WASP-69b, we investigated how
the photoevaporative outflow and \ext{He} observables depend
on various input parameters. 3D simulations are crucial for
capturing the full hydrodynamics including Coriolis force
and advection. These effects are needed for producing
correct line profile including the line ratios, kinematic
broadening and the overall blueshift. We found that EUV
photons are most efficient in driving the photoevaporation
dynamics and in producing \pos{He} as the progenitors of
recombination excitation of \ext{He}. The soft FUV photons
that can ionize \ext{He} often play an adverse effect on the
\ext{He} observability. X-ray photons, having much lower
interaction cross section, are of secondary
importance. Surface gravity also determines the
effectiveness of photoevaporative outflows with puffier
planets experiencing significantly stronger outflows, but
the response of He* equivalent width is sub-linear.

K-stars are at a sweet spot of FUV/EUV balance that maximize
the detectability of \ext{He} lines. F or earlier type stars
have excessive FUV fluxes that suppresses the \ext{He} lines
by orders of magnitude. G and M dwarfs represent a middle
ground: \ext{He} lines should still be observable
particularly for the younger and more active ones. In any
case, the depth \ext{He} line profiles cannot be translated
to a mass-loss rate without knowing the host star high
energy SED.

We also investigated whether stellar flares could explain
some of the temporal variability of WASP-69b
\citep{2018Sci...362.1388N}. We found that a flare which
enhances all high energy radiations initially suppresses
\ext{He} lines due to FUV fluxes ionizing the \ext{He}
before the whole system can adjust to higher mass-loss state
after some dynamical timescales (usually
hour-timescale). This characteristic shape is not consistent
with the observed temporal variability of WASP-69b which
only shows a decline of \ext{He} line depth before returing
to nominal levels. Stellar flares are unlikely to be the
explanation for this type of variability.

\vspace*{20pt} 

This work is supported by the Center for Computational
Astrophysics of the Flatiron Institute, and the Division of
Geological and Planetary Sciences of the California
Institute of Technology. L. Wang acknowledges the computing
resources provided by the Simons Foundation and the San
Diego Supercomputer Center. We thank our colleagues
(alphabetical order): Philip Armitage, Zhuo Chen, Jeremy
Goodman, Xiao Hu, Heather Knutson, Mordecai Mac-Low, Jessica
Spake, Kengo Tomida, Songhu Wang, Andrew Youdin, and Michael
Zhang, for helpful discussions and comments. We especially
thank Shreyas Vissapragada for detailed suggestions and
discussions.

\appendix

\section{Cores and Interal Atmospheres of Model Planets}
\label{sec:appdx-internal}

Gas giants like WASP-69b may have degenerate hydrogen and
helium in their interior. We adopt the equations of state
(EoS hereafter) tabulated by \citet{2016A&A...596A.114M},
which describe the behaviors of hydrogen and helium over
wide ranges of pressure and temperature, from the degenerate
states to ideal gases. Those tabulated EoS present the
density and entropy of hydrogen and helium as functions of
temperature and density.  The combined EoS for a mixture of
hydrogen and helium at a fixed atom number fraction
$x_{\chem{H}}$ is given by solving the equation for
$x_{p,\chem{H}}$ (the partial pressure of hydrogen),
\begin{equation}
  \label{eq:core-eos}
  \dfrac{m_{\chem{H}}x_{\chem{H}}}{m_{\chem{He}}(1-x_{\chem{H}})}
  = \dfrac{\rho_{\chem{H}}(p~x_{p,\chem{H}},T)}
  {\rho_{\chem{He}}[p (1 - x_{p,\chem{H}}),T]}\ ,
\end{equation}
in which $p$ is the total pressure, $T$ is the temperature,
$m_{\chem{H}}$ and $m_{\chem{He}}$ are the atomic masses of
hydrogen and helium respectively, and
$\rho_{\chem{H}},\ \rho_{\chem{He}}$ are interpolated from
the EoS tables. The overall density then reads
$\rho = \rho_{\chem{H}}+\rho_{\chem{He}}$. The entropy
density $s$ of the materials is also calculated, so that we
can obtain the adiabatic gradient,
\begin{equation}
  \label{eq:core-grad-ad}
  \nabla_\ad\equiv
  \left(\dfrac{\partial\ln p}{\partial \ln T}\right)_s
  = -\dfrac{(\partial \ln s/\partial \ln p)_T}
  {(\partial \ln s/\partial \ln T)_p}\ .
\end{equation}
Note that the entropy of mixing does not affect these
derivatives.

The spherical symmetric isentropic hydrostatics are
calculated by solving a boundary value problem for the set
of ordinary differential equations:
\begin{equation}
  \label{eq:core-hydro}
    \dfrac{\d p}{\d r} = - \dfrac{G M \rho}{r^2}\ ,
    \quad \dfrac{\d M}{\d r} = 4\pi r^2 \rho\ ,
    \quad \dfrac{\d T}{\d r}
    = \nabla_\ad\left(\dfrac{T}{p}\right)
    \left( \dfrac{\d p}{\d r} \right)\ .
\end{equation}
where $M$ denotes the mass enclosed by radius $r$.
Specifying the boundary conditions $(p_c,\ T_c)$ as the
``eigenvalues'', we can integrate these ODEs from a the
boundary of a dense solid core with radius $r_c$ and given
mass $M_c$ up to the radiative-convective boundary
$r = r_\rcb$. At $r_\rcb$ the temperature approaches the
equilibrium temperature of the quasi-isothermal layer
$T = T_\eq$. Thus the convective inner atmosphere is
smoothly connected to an quasi-isothermal outer atmosphere
whose density profile obeys ($\mu$ is the dimensional mean
molecular mass),
\begin{equation}
  \label{eq:core-isothermal}
  \rho = \rho_\rcb \exp\left[
    \left(\dfrac{G M_\rcb\mu}{r_\rcb \kb T_\eq}\right)
    \left( \dfrac{r_\rcb}{r} - 1 \right) \right]\ .
\end{equation}
The density profile is then used to calculate the effective
transiting radius,
\begin{equation}
  \label{eq:core-transiting}
  \mean{r_\eff} = \left\{ \dfrac{1}{\pi} \int_0^{r_{\rm cut}}
    \d b\ 2\pi b\left[1-\e^{-\tau(b)}\right] \right\}^{1/2}\ ,
\end{equation}
where $r_{\rm cut}$ is an arbitrary cutoff size (to
calculate the effective transiting radius in the broad
optical band, we use $r_{\rm cut} = 100~R_\oplus$),
$\tau(b)$ is the optical depth along the LoS at impact
parameter $b$,
\begin{equation}
  \label{eq:core-tau-b}  
  \tau(b) \equiv \int_{-r_{\rm cut}}^{r_{\rm cut}}\d z
    \ \sigma n|_{r=\sqrt{b^2+z^2}}\ , 
\end{equation}
$n$ is the number density of the extinction particle, and
$\sigma$ is the extinction cross section per particle. The
eigenvalues $(p_c,\ T_c)$ are searched iteratively until
both $M_\rcb$ and $\mean{r_\eff}$ match the observed the
mass $M_\p$ and optical transiting radius $R_\p$ of the
planet being simulated. In all models discussed in this
paper, for simplicity, we assume that there is no rocky
cores $r_c = 0$, $M_c = 0$. This assumption hardly affects
the properties of the upper atmosphere.  We also use
$\kappa$ of the Thomson scattering to estimate calculate
$\mean{r_\eff}$. We have also tested other plausible values
of opacity (e.g. the optical $\kappa$ for $r_\dust = 5~\ang$
very small grains with $10^{-4}$ dust-to-gas mass ratio),
and the $\mean{r_\eff}$ varies by only $\sim 2\%$ under the
same boundary conditions. Again the specific choice of
opacity hardly affects the upper atmosphere.

\bibliography{planet_he.bib}
\bibliographystyle{aasjournal}

%
\end{document}